%% file: main.tex
\documentclass[conference]{IEEEtran}

\input{include/package.tex}
\input{include/command.tex}
\input{include/shorthand.tex}

\begin{document}

% Title
\title{TACOS: Topology-Aware Collective Algorithm Synthesizer for Distributed Machine Learning}
\input{include/authors.tex}
\maketitle

% Abstract
\input{content/0_abstract.tex}

% Keywords
\begin{IEEEkeywords}
collective communication, collective algorithm synthesizer, distributed machine learning
\end{IEEEkeywords}

% Content
\input{content/1_introduction.tex}
\input{content/2_background.tex}
\input{content/3_motivation.tex}
\input{content/4_tacos.tex}
\input{content/5_methodology.tex}
\input{content/6_result.tex}
\input{content/7_related_work.tex}
\input{content/8_conclusion.tex}

% Acknowledgment
\input{include/acknowledgment.tex}

% Artifact Appendices
\input{content/9_artifact.tex}

% References
\bibliographystyle{IEEEtran}
\bibliography{reference/reference.bib}

% End of document
\end{document}

%% file: include/package.tex
\usepackage{cite}
\usepackage{amsmath,amssymb,amsfonts}
\usepackage{graphicx}
\usepackage{textcomp}
\usepackage{xcolor}
\def\BibTeX{{\rm B\kern-.05em{\sc i\kern-.025em b}\kern-.08em
    T\kern-.1667em\lower.7ex\hbox{E}\kern-.125emX}}
    
\usepackage[normalem]{ulem}
\usepackage{soul}
\usepackage[export]{adjustbox}
\usepackage{multirow}
\usepackage{enumitem}
\usepackage{lipsum}
\setlipsum{
  par-before = \begingroup\color{gray},
  par-after = \endgroup
}
\usepackage{algorithm}
\usepackage[noend]{algpseudocode}
\usepackage{tcolorbox}

\usepackage[bookmarks=true,breaklinks=true,letterpaper=true,colorlinks,citecolor=blue,linkcolor=blue,urlcolor=blue]{hyperref}

%% file: include/command.tex
\newcommand{\insertFigure}[5]{
    \begin{figure}[t]
      \centering
      \includegraphics[width=#3\linewidth]{figure/#1}
      \vspace{#4}
      \caption{\small #2}
      \label{fig:#1}
      \vspace{#5}
    \end{figure}
}

\newcommand{\insertFigureWide}[5]{
    \begin{figure*}[t]
      \centering
      \includegraphics[width=#3\linewidth]{figure/#1}
      \vspace{#4}
      \caption{\small #2}
      \label{fig:#1}
      \vspace{#5}
    \end{figure*}
}

%% file: include/shorthand.tex
\newcommand{\framework}{\textsc{Tacos}}

\newcommand{\ten}{TEN}

\newcommand{\reducescatter}{Reduce-Scatter}
\newcommand{\allgather}{All-Gather}
\newcommand{\allreduce}{All-Reduce}

\newcommand{\ring}{Ring}
\newcommand{\fullyconnected}{FullyConnected}
\newcommand{\mesh}{2D Mesh}
\newcommand{\hypercube}{3D Hypercube}
\newcommand{\switch}{Switch}

\newcommand{\torus}{3D Torus}

\newcommand{\dragonfly}{DragonFly}
\newcommand{\df}{DF}
\newcommand{\rfs}{3D-RFS}

\newcommand{\ringAlg}{Ring}
\newcommand{\direct}{Direct}
\newcommand{\hd}{RHD}
\newcommand{\halvingDoubling}{Recursive Halving-Doubling}
\newcommand{\dbt}{DBT}
\newcommand{\doubleTree}{Double Binary Tree}
\newcommand{\hierarchical}{BlueConnect}
\newcommand{\themis}{Themis}
\newcommand{\ccube}{C-Cube}
\newcommand{\multitree}{MultiTree}
\newcommand{\blink}{Blink}
\newcommand{\sccl}{SCCL}
\newcommand{\taccl}{TACCL}
\newcommand{\tto}{TTO}

\newcommand{\gnmt}{GNMT}
\newcommand{\resnet}{ResNet-50}
\newcommand{\tnlg}{Turing-NLG}
\newcommand{\msft}{MSFT-1T}

%% file: include/authors.tex
\author{

\IEEEauthorblockN{
    William Won\IEEEauthorrefmark{1},
    Midhilesh Elavazhagan\IEEEauthorrefmark{2},
    Sudarshan Srinivasan\IEEEauthorrefmark{2},
    Swati Gupta\IEEEauthorrefmark{3},
    and Tushar Krishna\IEEEauthorrefmark{1}
}
\IEEEauthorblockA{
    \IEEEauthorrefmark{1}Georgia Institute of Technology\ \ \ 
    \IEEEauthorrefmark{2}Intel\ \ \  
    \IEEEauthorrefmark{3}Massachusetts Institute of Technology
}
\IEEEauthorblockA{
    \IEEEauthorrefmark{1}william.won@gatech.edu\ \ \ 
    \IEEEauthorrefmark{2}midhilesh.elavazhagan@intel.com\ \ \ 
    \IEEEauthorrefmark{2}sudarshan.srinivasan@intel.com\\
    \IEEEauthorrefmark{3}swatig@mit.edu\ \ \ 
    \IEEEauthorrefmark{1}tushar@ece.gatech.edu
}

\vspace{-2em}
}

%% file: content/0_abstract.tex
\begin{abstract}

The surge of artificial intelligence, particularly large language models, has driven the rapid development of large-scale machine learning clusters.
Executing distributed models on these clusters is often constrained by communication overhead, making efficient utilization of available network resources crucial.
As a result, the routing algorithm employed for collective communications (i.e., collective algorithms) plays a pivotal role in determining overall performance.
Unfortunately, existing collective communication libraries for distributed machine learning are limited by a fixed set of basic collective algorithms.
This limitation hinders communication optimization, especially in modern clusters with heterogeneous and asymmetric topologies.
Furthermore, manually designing collective algorithms for all possible combinations of network topologies and collective patterns requires heavy engineering and validation efforts.
To address these challenges, this paper presents \framework, an autonomous synthesizer capable of automatically generating topology-aware collective algorithms tailored to specific collective patterns and network topologies.
\framework\ is highly flexible, synthesizing an \allreduce\ algorithm for a heterogeneous 128-NPU system in just 1.08 seconds, while achieving up to a 4.27$\times$ performance improvement over state-of-the-art synthesizers.
Additionally, \framework\ demonstrates better scalability with polynomial synthesis times, in contrast to NP-hard approaches which only scale to systems with tens of NPUs.
\framework\ can synthesize for 40K NPUs in just 2.52 hours.

\end{abstract}

%% file: content/1_introduction.tex
\section{Introduction}\label{sec:intro}

The advancement of Artificial Intelligence (AI) has led to a recent surge in the design of specialized distributed High-performance Computing (HPC) platforms tailored for AI, often called AI supercomputers. 
Examples include Intel Gaudi~\cite{HabanaPtP}, Google Cloud TPU~\cite{cloudTpuPaper}, NVIDIA HGX~\cite{NvidiaH100}, Cerebras Andromeda~\cite{CerebrasCS2}, Tesla Dojo~\cite{TeslaDojo}, and many more. 
From a bird's eye view, these platforms incorporate multiple Neural Processing Units (NPUs, such as GPUs, TPUs, or ASICs) at the endpoints, interconnected by custom high-speed fabrics. 
These AI supercomputers are of particular importance for enormous Deep Neural Network (DNN) models, such as Large Language Models (LLMs), due to their substantial compute and memory demands~\cite{mszero}. 

It is important to note that different AI clusters employ diverse network topologies. 
For instance, point-to-point connections are used in Intel Gaudi~\cite{HabanaPtP} and Google TPU~\cite{cloudTpuPaper}, hierarchical switches are utilized in NVIDIA HGX~\cite{NVIDIASuperPod}, mesh architectures are employed in Tesla Dojo~\cite{TeslaDojo} and Cerebras Andromeda~\cite{CerebrasCS2}, and DragonFly~\cite{dragonfly} connections are used in Groq~\cite{groq}. 
Moreover, these systems often incorporate multiple link technologies such as XeLink~\cite{xelink}, NVLink~\cite{nvlinkBridge}, InfiniBand~\cite{infiniband}, or even optical networks~\cite{cloudTpuPaper}. 
In summary, topologies in Machine Learning (ML) clusters commonly exhibit \emph{asymmetric}\footnote{ 
For e.g., NPUs at the center vs. edge of a \mesh\ have different degrees. 
} shapes and \emph{heterogeneous}\footnote{ 
All links within a \emph{homogeneous} topology have the same bandwidth. 
} link bandwidths. 
Considering that communication among NPUs is the main bottleneck of distributed ML~\cite{CommBottleneck1,CommBottleneck2,CommBottleneck3}, orchestrating communication within these AI platforms remains an open research problem~\cite{bottleneck1, bottleneck2,themis,taccl,sccl}.

\input{table/compare_collective_algs.tex}

\input{table/compare_tools.tex}

\insertFigureWide{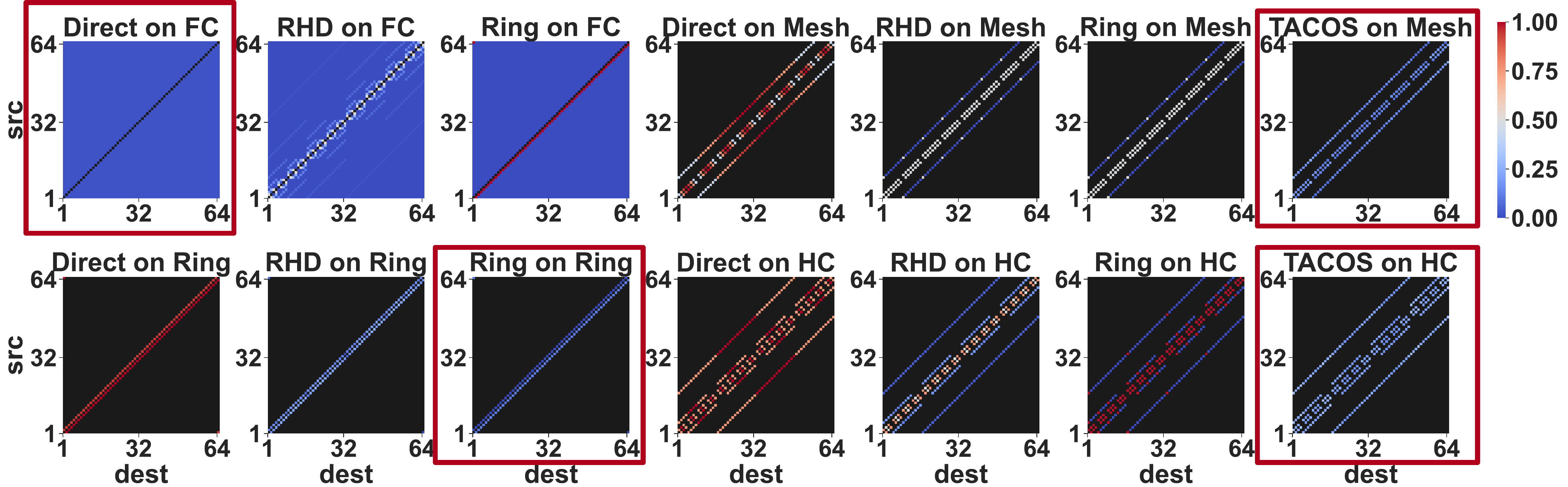}{
Heat map of total message size transferred over each link, when running 1 GB \allreduce\ using basic algorithms (\direct, \hd, \ringAlg, and \framework\ for comparison) over different network topologies (\fullyconnected~(FC), Ring, \mesh, and \hypercube~(HC)). 
Each cell at (\emph{src}, \emph{dest}) denotes a link connecting NPU \emph{src} to NPU \emph{dest}. 
If there is no such link in the topology, the cell is marked black. 
All values are normalized to the largest value per topology. 
For every topology, scenarios running topology-aware collective algorithms are marked with a red box, which results in balanced, lowest overall loads (i.e., a cooler heat map).
}{1}{-2em}{-1em}

Communication within AI workloads presents unique characteristics compared to traditional cloud workloads. 
In the latter, there can be heavy variability in the traffic pattern and volumes. 
This has naturally led to a set of past works on dynamically managing network congestion in data center fabrics~\cite{saeedHoti}. 
In contrast, for ML models, both the traffic pattern and traffic volume are \emph{deterministic} once the workload has been partitioned and mapped~\cite{deepspeed}. 
The traffic pattern is \emph{collective} in nature (i.e., one-to-many and many-to-one). 
Given these characteristics, AI workload communication today is often routed \emph{statically} via Collective Communication Libraries (CCLs)~\cite{nccl,rccl,msccl,oneCCLDoc}. 
These CCLs encompass a collection of predefined collective algorithms known as \emph{topology-aware collectives}. 
As the name suggests, these algorithms are tailored for operation on specific network topologies~\cite{collectiveNetworkDependent}. 
For example, \ringAlg\ (RI)~\cite{ringCollective}, \direct\ (DI)~\cite{astrasim}, \halvingDoubling\ (\hd)~\cite{rabenseifner}, and \doubleTree\ (\dbt)~\cite{dbt} are distinct \allreduce\ collective algorithms tailored for fundamental network topologies (a.k.a. basic collective algorithms). 
\autoref{table:compare_collective_algs} exemplifies several predefined collective algorithms and their target topologies.

\insertFigure{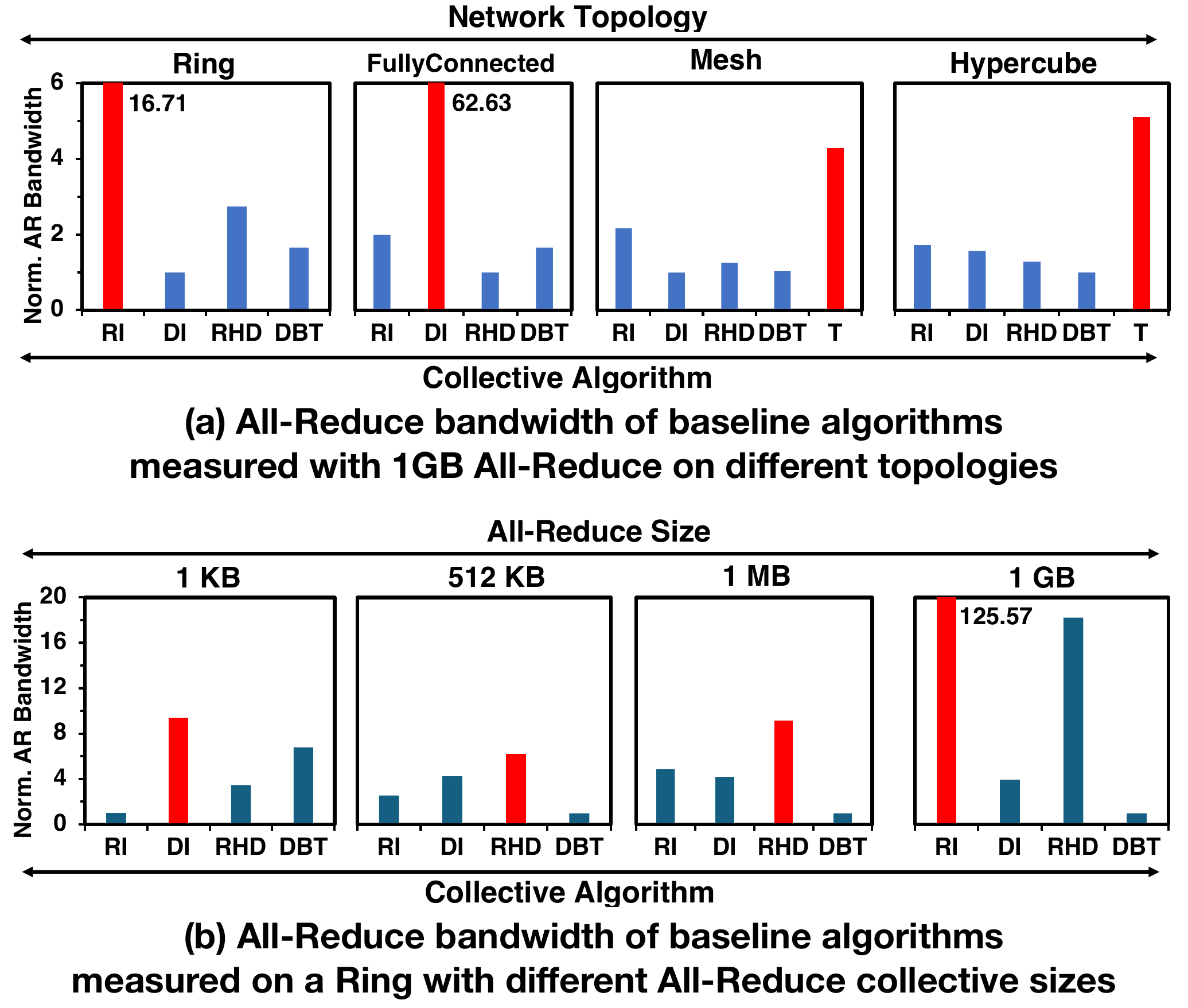}{
(a)~\allreduce\ bandwidth~(i.e., collective size $\div$ collective time) over distinct network topologies with 64 NPUs (using link $\alpha$=0.5µs, 1/$\beta$=50GB/s, as explained in \autoref{subsec:tacos_extension}), measured with 1 GB collective. 
For \mesh\ and \hypercube, we also run the topology-aware algorithm synthesized by \framework~(T). 
(b)~\allreduce\ bandwidth for different collective algorithms on a 128-NPU physical \ring\ (link $\alpha$=30ns, 1/$\beta$=150GB/s), measured with varying collective sizes. 
Results are normalized per each graph by the smallest number.
}{1}{-2em}{-1em}

\insertFigureWide{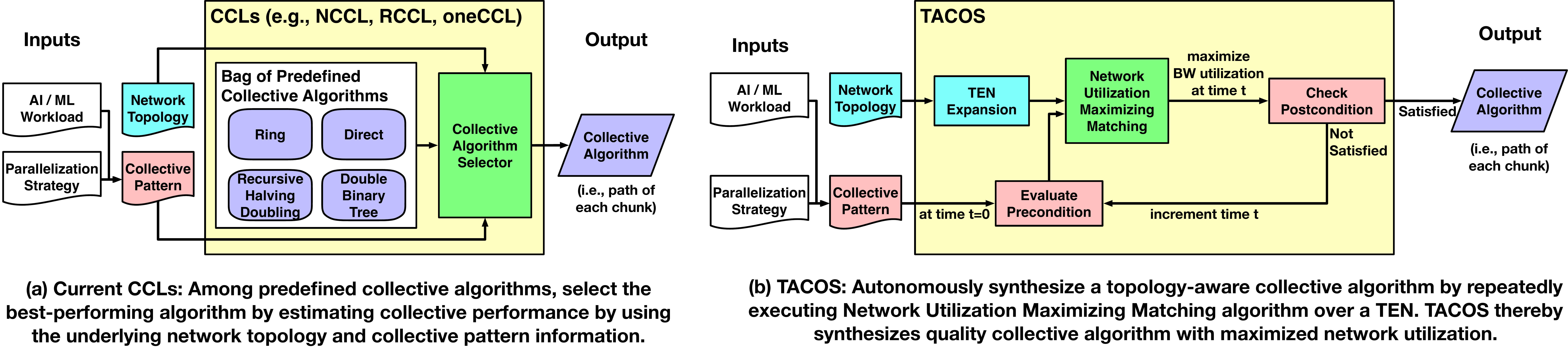}{
(a)~Current CCLs execute collectives by selecting an algorithm from a set of predefined implementations. 
(b)~High-level overview of the \framework\ framework. 
The target network topology and collective pattern are provided as inputs. 
\framework\ expands the network into a \ten\ and evaluates the collective precondition (i.e., which chunks are currently held by each NPU) for time $t$=0. 
Based on this information, \framework\ employs a chunk-link matching algorithm that maximizes network resource utilization for the target time $t$=0. 
\framework\ iteratively runs this matching process for successive time spans until the postcondition is satisfied (i.e., all NPUs have received every desired chunk). 
The details are explained in \autoref{sec:greedyBased}. 
This procedure yields a topology-aware collective algorithm (i.e., static path of each chunk), which can then be utilized by CCLs in lieu of the predefined topology-unaware basic algorithms.
}{1}{-2em}{-1em}

Naively deploying topology-aware collectives over non-preferred physical topologies could induce link congestions and underutilization of the network resources.
This is demonstrated in \autoref{fig:ResultMotivationHeatmap}.
When basic collective algorithms are executed over their non-preferred topologies, significant link underutilization and oversubscriptions occur.\footnote{
We use \emph{bidirectional} Ring algorithm and \emph{bidirectional} Ring topology for all the motivation and evaluation data in this paper.
} This is the reason CCLs today have heuristics built-in to switch among basic collective algorithms.
Unfortunately, beyond a limited set of basic collective algorithms, there are no known solutions for arbitrary topologies.
Given this complexity, when given an unknown topology, CCLs today default to employing the \ring\ algorithm.\footnote{
Either one logical ring is mapped over the physical topology, or multiple parallel rings~\cite{blueconnect,themis}.
But the problem of over/undersubscription remains.
} However, this approach may result in oversubscription and underutilization of certain links.

To address the aforementioned challenge, recent studies have demonstrated the feasibility and effectiveness of \emph{autonomously synthesizing} topology-aware collective algorithms when provided with a target physical network and collective patterns~\cite{blink, sccl, multitree, taccl}.
These efforts are outlined in \autoref{table:compare_tools}.
However, they are limited to supporting a specific subset of topologies, typically homogeneous or symmetric networks.
Furthermore, they often overlook network congestion effects during the search, resulting in suboptimal synthesis results.
A number of works treat collective synthesis as a global optimization problem, which poses NP-hard complexity to the search space, fundamentally limiting their scalability~\cite{ilpNPHard}.
Consequently, their evaluations have been restricted to only tens of NPUs.

In this work, we propose an autonomous collective synthesizer named \textbf{\framework: \uline{T}opology-\uline{A}ware \uline{Co}llective Algorithm \uline{S}ynthesizer}.\footnote{
\framework\ is open-sourced at: \url{https://github.com/astra-sim/tacos}.
}
\framework\ architecture is summarized in \autoref{fig:TACOSArchitecture}. 
Given a network topology and target collective pattern, \framework\ autonomously synthesizes a \emph{static} collective algorithm, which can be leveraged by CCLs. 
This is done by \emph{iteratively maximizing network resource utilization} throughout the course of collective execution. 
\framework\ showcases better quality synthesis results, orders of magnitude better scalability, and encompasses a larger class of topologies compared to previous endeavors. 
This is enabled by two features. 
Firstly, we frame collective algorithm synthesis as a \textit{link-chunk matching problem} using a Time-expanded Network (\ten) representation, an acyclic directed graph integrating spatial and temporal dimensions~\cite{tenRepr1, tenUse2, tenDefinition}. 
Secondly, we employ a \textit{link-chunk matching algorithm} to maximize network utilization and minimize congestion. Here is a summary of our contributions:

\input{table/ps_collectives.tex}

\begin{itemize}[leftmargin=*]
 \item \framework\ brings the TEN representation into the domain of distributed ML, which is a widely adopted data structure for traffic and flow optimizations.
 \item \framework\ accommodates a wide range of arbitrary (i.e., heterogeneous and asymmetric) topologies.
 \item \framework\ considers network congestion effects throughout the synthesis process, resulting in high-quality search outcomes.
 \item \framework\ features better scalability, surpassing previous works usually restricted to tens of NPUs. \framework\ synthesized collective algorithms for a 40K NPU topology in 2.52 hours.
\end{itemize}

%% file: table/compare_collective_algs.tex
\begin{scriptsize}
\begin{table}[t!]
\centering
\caption{
\allreduce\ algorithms and their target topologies. Topology abbreviations denote RI~(\ring), FC~(\fullyconnected), SW~(\switch), Ho~(Homogeneous), and Ht~(Heterogeneous). \\
$\triangle$ denotes the collective performance is closely tied to specific network configurations.
}
\label{table:compare_collective_algs}
\vspace{-0.5em}

\begin{tabular}{|c|cccccccc|}
\hline
\multirow{3}{*}{\textbf{\begin{tabular}[c]{@{}c@{}}Topology-aware\\ \allreduce\\ Algorithm\end{tabular}}} & \multicolumn{8}{c|}{\textbf{Preferred Physical Topology}} \\ \cline{2-9} 
 & \multicolumn{3}{c|}{\textbf{Uni-dimensional}} & \multicolumn{2}{c|}{\textbf{\begin{tabular}[c]{@{}c@{}}Multi-\\ dim.\end{tabular}}} & \multicolumn{2}{c|}{\textbf{\begin{tabular}[c]{@{}c@{}}Asym-\\ metric\end{tabular}}} & \textbf{Any} \\ \cline{2-9} 
 & \multicolumn{1}{c|}{\textbf{RI}} & \multicolumn{1}{c|}{\textbf{FC}} & \multicolumn{1}{c|}{\textbf{SW}} & \multicolumn{1}{c|}{\textbf{Ho}} & \multicolumn{1}{c|}{\textbf{Ht}} & \multicolumn{1}{c|}{\textbf{Ho}} & \multicolumn{1}{c|}{\textbf{Ht}} & \textbf{Any} \\ \hline
\textbf{\ring}~\cite{ringCollective} & \multicolumn{1}{c|}{$\checkmark$} & \multicolumn{1}{c|}{} & \multicolumn{1}{c|}{} & \multicolumn{1}{c|}{} & \multicolumn{1}{c|}{} & \multicolumn{1}{c|}{} & \multicolumn{1}{c|}{} &  \\ \hline
\textbf{\direct}~\cite{astrasim} & \multicolumn{1}{c|}{} & \multicolumn{1}{c|}{$\checkmark$} & \multicolumn{1}{c|}{} & \multicolumn{1}{c|}{} & \multicolumn{1}{c|}{} & \multicolumn{1}{c|}{} & \multicolumn{1}{c|}{} &  \\ \hline
\textbf{RHD}~\cite{rabenseifner} & \multicolumn{1}{c|}{} & \multicolumn{1}{c|}{} & \multicolumn{1}{c|}{$\checkmark$} & \multicolumn{1}{c|}{} & \multicolumn{1}{c|}{} & \multicolumn{1}{c|}{} & \multicolumn{1}{c|}{} &  \\ \hline
\textbf{DBT}~\cite{dbt} & \multicolumn{1}{c|}{} & \multicolumn{1}{c|}{} & \multicolumn{1}{c|}{$\checkmark$} & \multicolumn{1}{c|}{} & \multicolumn{1}{c|}{} & \multicolumn{1}{c|}{} & \multicolumn{1}{c|}{} &  \\ \hline
\textbf{\begin{tabular}[c]{@{}c@{}}\hierarchical\\ ~\cite{blueconnect}\end{tabular}} & \multicolumn{1}{c|}{$\checkmark$} & \multicolumn{1}{c|}{$\checkmark$} & \multicolumn{1}{c|}{$\checkmark$} & \multicolumn{1}{c|}{$\triangle$} & \multicolumn{1}{c|}{$\triangle$} & \multicolumn{1}{c|}{} & \multicolumn{1}{c|}{} &  \\ \hline
\textbf{\themis}~\cite{themis} & \multicolumn{1}{c|}{$\checkmark$} & \multicolumn{1}{c|}{$\checkmark$} & \multicolumn{1}{c|}{$\checkmark$} & \multicolumn{1}{c|}{$\checkmark$} & \multicolumn{1}{c|}{$\checkmark$} & \multicolumn{1}{c|}{} & \multicolumn{1}{c|}{} &  \\ \hline
\textbf{\tto}~\cite{tto} & \multicolumn{1}{c|}{} & \multicolumn{1}{c|}{} & \multicolumn{1}{c|}{} & \multicolumn{1}{c|}{$\triangle$} & \multicolumn{1}{c|}{} & \multicolumn{1}{c|}{$\triangle$} & \multicolumn{1}{c|}{} &  \\ \hline
\textbf{\ccube}~\cite{ccube} & \multicolumn{1}{c|}{} & \multicolumn{1}{c|}{$\triangle$} & \multicolumn{1}{c|}{} & \multicolumn{1}{c|}{$\triangle$} & \multicolumn{1}{c|}{} & \multicolumn{1}{c|}{$\triangle$} & \multicolumn{1}{c|}{} &  \\ \hline
\textbf{\begin{tabular}[c]{@{}c@{}}\framework\\ (this work)\end{tabular}} & \multicolumn{1}{c|}{$\checkmark$} & \multicolumn{1}{c|}{$\checkmark$} & \multicolumn{1}{c|}{$\checkmark$} & \multicolumn{1}{c|}{$\checkmark$} & \multicolumn{1}{c|}{$\checkmark$} & \multicolumn{1}{c|}{$\checkmark$} & \multicolumn{1}{c|}{$\checkmark$} & $\checkmark$ \\ \hline
\end{tabular}

\vspace{-0.5em}

\end{table}
\end{scriptsize}

%% file: table/compare_tools.tex
\begin{scriptsize}
\begin{table}[t!]
\centering
\caption{
Qualitative comparison of collective algorithm synthesizers.  
\taccl\ is marked $\triangle$ as it requires assumptions (communication sketch) to model network heterogeneity and congestion.
}
\label{table:compare_tools}
\vspace{-0.5em}

\begin{tabular}{|c|cc|ccc|}
\hline
\multirow{2}{*}{\textbf{Framework}} & \multicolumn{2}{c|}{\textbf{Network}} & \multicolumn{3}{c|}{\textbf{Synthesis Mechanism}} \\ \cline{2-6} 
 & \multicolumn{1}{c|}{\textbf{\begin{tabular}[c]{@{}c@{}}Asym-\\ metric\end{tabular}}} & \textbf{\begin{tabular}[c]{@{}c@{}}Hetero-\\ geneous\end{tabular}} & \multicolumn{1}{c|}{\textbf{\begin{tabular}[c]{@{}c@{}}Auton-\\ omous\end{tabular}}} & \multicolumn{1}{c|}{\textbf{\begin{tabular}[c]{@{}c@{}}Removes\\ Conges-\\ tion\end{tabular}}} & \textbf{\begin{tabular}[c]{@{}c@{}}Scal-\\ able\end{tabular}} \\ \hline
\textbf{\sccl}~\cite{sccl} & \multicolumn{1}{c|}{} &  & \multicolumn{1}{c|}{$\checkmark$} & \multicolumn{1}{c|}{} &  \\ \hline
\textbf{\blink}~\cite{blink} & \multicolumn{1}{c|}{$\checkmark$} &  & \multicolumn{1}{c|}{$\checkmark$} & \multicolumn{1}{c|}{} &  \\ \hline
\textbf{\multitree}~\cite{multitree} & \multicolumn{1}{c|}{$\checkmark$} &  & \multicolumn{1}{c|}{$\checkmark$} & \multicolumn{1}{c|}{$\checkmark$} &  \\ \hline
\textbf{\taccl}~\cite{taccl} & \multicolumn{1}{c|}{} & $\triangle$ & \multicolumn{1}{c|}{$\triangle$} & \multicolumn{1}{c|}{$\triangle$} &  \\ \hline
\textbf{\begin{tabular}[c]{@{}c@{}}\framework\\ (this work)\end{tabular}} & \multicolumn{1}{c|}{$\checkmark$} & $\checkmark$ & \multicolumn{1}{c|}{$\checkmark$} & \multicolumn{1}{c|}{$\checkmark$} & $\checkmark$ \\ \hline
\end{tabular}

\vspace{-1em}
\end{table}
\end{scriptsize}

%% file: table/ps_collectives.tex
\begin{scriptsize}
\begin{table}[t]
\centering
\caption{Common model and data parallelization strategies and their required collective communication patterns.
}
\label{table:psCollectives}
\vspace{-0.5em}

\begin{tabular}{|c|c|c|c|}
\hline
\textbf{Parallelization} & \textbf{\reducescatter} & \textbf{\allgather} & \textbf{\allreduce} \\ \hline
Data Parallelism~\cite{astrasim} &  &  & $\checkmark$ \\ \hline
Tensor Parallelism~\cite{astrasim} &  &  & $\checkmark$ \\ \hline
FSDP~\cite{fsdp} & $\checkmark$ & $\checkmark$ &  \\ \hline
ZeRO~\cite{mszero} & $\checkmark$ & $\checkmark$ &  \\ \hline
Hybrid~\cite{astrasim} & $\checkmark$ & $\checkmark$ & $\checkmark$ \\ \hline
\end{tabular}
\vspace{-2em}

\end{table}

\end{scriptsize}

%% file: content/2_background.tex
\section{Background}\label{sec:Background}

\subsection{Collective Communication Patterns}\label{subsec:BasicCollective}

Communication poses a significant challenge for distributed ML systems due to the dispersal of models (and training data for training tasks) across devices, requiring frequent synchronization among NPUs~\cite{themis}.  
These communications can be represented as collective communications~\cite{astrasim}.  
As depicted in \autoref{table:psCollectives}, workload parallelization strategies require specific collective patterns to be executed.  
\autoref{fig:CollectiveDefinition} illustrates common collective patterns.  
Each circle represents a chunk, the atomic unit for scheduling collectives across the network.  
The \allreduce\ can be conceptualized as two sequential stages: \reducescatter\ followed by \allgather, and it is the most prevalent pattern in synchronous NPU-to-NPU ML executions~\cite{rinective}.  
To enhance network utilization, a collective can be decomposed into multiple smaller chunks which can be run in parallel~\cite{themis}.

\insertFigure{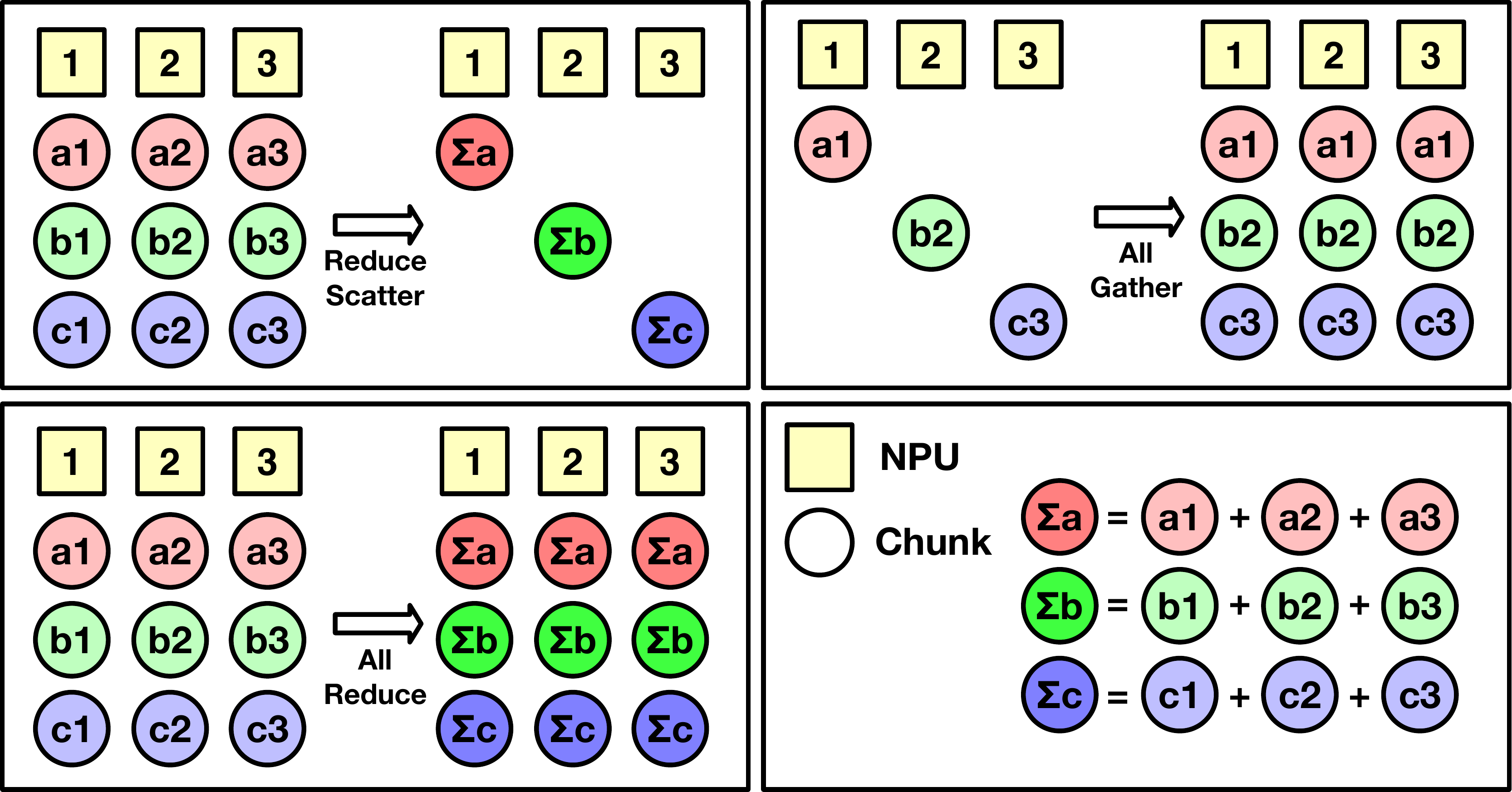}{
Common collective patterns used in distributed ML.  
Each collective communication defines a specific data transfer pattern.  
For example, in \allgather, each NPU starts with one chunk (denoted as a circle) and broadcasts it to all other NPUs.  
In \reducescatter, all NPUs start with $N$ chunks and end up with one chunk, constructed by summing the corresponding chunks from all other NPUs.
}{1}{-1.8em}{-1em}

\subsection{Collective Communication Algorithms}

For a given collective communication pattern, there can be several distinct routing algorithms to implement it.  
These are called \textit{collective algorithms} and define the \emph{static path for each chunk during the execution of collective communication}.  
Example collective algorithms' traffic patterns are depicted in~\autoref{fig:CollectiveAlgorithmPattern}.  
The intuition for having different algorithms for the same collective pattern is to minimize contention on distinct physical network topologies.

%% file: content/3_motivation.tex
\section{Motivation and Problem Statement}

\subsection{Importance of Topology-aware Collectives}\label{sec:importance_tac}

The physical characteristics of the network topology (including connectivity, link latencies, and link bandwidths) heavily influence network transmission behavior.  
We define a \emph{topology-aware collective algorithm} as one intricately optimized alongside the underlying topology to achieve optimal collective performance~\cite{topologyAwareColl,rabenseifner}.  
\autoref{table:compare_collective_algs} summarizes common topology-aware algorithms and their target topologies.

Example traffic patterns of basic collective algorithms are depicted in \autoref{fig:CollectiveAlgorithmPattern}.  
When these patterns are mapped over their \emph{preferred} topologies, they can efficiently utilize links in the topology every cycle without any contention.  
However, executing a collective algorithm on its unpreferred physical topology leads to contention (i.e., oversubscription) on certain links and low or even no utilization (i.e., undersubscription) on others.  
This effect can be observed in the link subscription heat map in \autoref{fig:ResultMotivationHeatmap}.  
We also quantify these effects in \autoref{fig:ResultMotivationMeasurement}(a).  
On the \ring\ network, the topology-aware \ringAlg\ algorithm exhibited a $16.71\times$ higher \allreduce\ bandwidth than the \direct\ algorithm.  
Conversely, on the \fullyconnected\ topology, the \direct\ algorithm demonstrated a performance improvement of $62.63\times$ over the \ringAlg\ algorithm.  
\autoref{fig:ResultMotivationMeasurement}(b) also demonstrates that the optimal topology-aware collective could change depending on the target collective pattern.  
Unlike the 1 GB \allreduce\ case, for 1 KB, the \direct\ algorithm outperformed \ringAlg\ since each network transmission is latency-bound and prefers short-distance collective algorithms.  

These results emphasize the importance of designing optimal topology-aware collective algorithms that encapsulate both communication patterns and network topology.  
Modern CCLs~\cite{nccl, oneCCLDoc, rccl} implement a range of predefined basic collective algorithms, selecting them based on network features as shown in \autoref{fig:TACOSArchitecture}(a).  
Unfortunately, this fundamentally limits the physical topologies they can efficiently support.

\subsection{Heterogeneity, Asymmetry, and Scale of ML Systems}\label{sec:heterogeneity_symmetry}

To provide maximum network resources, state-of-the-art distributed ML platforms are leveraging various network topologies and technologies, introducing bandwidth heterogeneity and asymmetry.  
Notable examples include asymmetric \mesh\ topology with wafer-scale technologies as implemented by Cerebras CS-2~\cite{CerebrasCS2} and Tesla Dojo~\cite{TeslaDojo} systems.  
Heterogeneous multi-dimensional scale-up and scale-out switches in NVIDIA DGX clusters~\cite{nvlinkBridge, infiniband}, and the integration of scale-up and photonic technologies by Google Cloud TPU~\cite{cloudTpuPaper, tpuv4} are other examples.  
Moreover, the scale of recent ML clusters is also increasing to accommodate larger workloads, incorporating thousands to even tens of thousands of NPUs.  
This trend is evident in Google Cloud TPUv5 pods with 8,960 TPUs~\cite{google2023tpupod}, the OpenAI cluster with 7,500 GPUs~\cite{sigler2021openaicluster}, and Meta's Research Supercluster equipped with 24K GPUs~\cite{meta2024rsc}.

\insertFigure{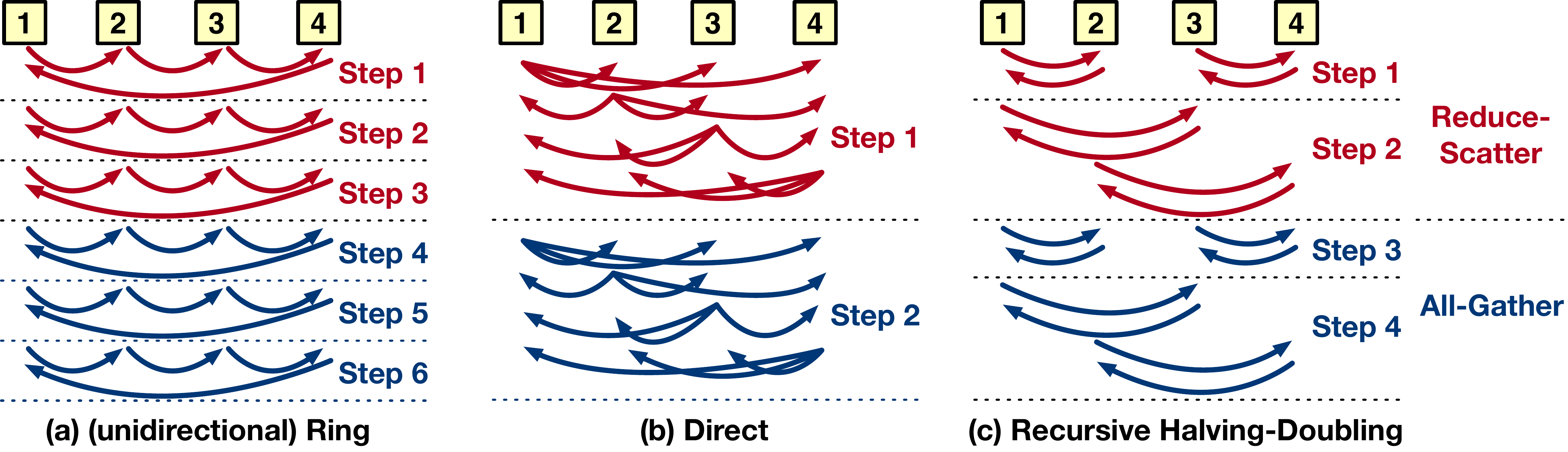}{
Example \allreduce\ algorithms and their traffic patterns.  
In \reducescatter, each red arrow represents sending and adding a chunk to the local data.  
For \allgather, each blue arrow signifies forwarding a chunk.  
Depending on the network connectivity, each step may encounter link congestion and underutilization.
}{1}{-2em}{0em}

\subsection{Challenge: Designing Topology-aware Collectives}\label{subsec:arbitraryAlgo}

Given the importance of topology-aware collective algorithms in distributed ML, designing them for commonly used networks is an active research field.  
For example, recent proposals devise \allreduce\ algorithms for \mesh~\cite{meshAlg} and \dragonfly~\cite{dragonflyAlg} topologies.  
However, this process requires engineering and validation efforts for each new topology variant.  
Flattened Butterfly~\cite{flatbutter}, MegaFly~\cite{megafly}, SlimFly~\cite{slimfly}, and Tofu~\cite{tofu} are just a few examples that do not yet have specialized collective algorithms and default to baseline collective algorithms, which could lead to network resource underutilization.  
Even existing collective algorithms are often tuned under multiple assumptions, such as homogeneous link bandwidths or specific message sizes, making them susceptible to potential performance degradation in practice.

\subsection{Solution: Topology-aware Collective Synthesizer}

To address the challenges inherent in designing topology-aware collective algorithms manually, the development of an \emph{autonomous synthesis framework capable of automatically generating topology-aware collective algorithms} is imperative.  
Such a toolchain can alleviate the burden on human experts to manually configure the routing of each chunk.
Additionally, by automatically exploring the design space, these frameworks have the potential to generate higher-quality collective algorithms for specific targets.
This is particularly crucial as network topologies become larger and more intricate, further complicating the task of design and optimization.

%% file: content/4_tacos.tex
\section{TACOS}\label{sec:greedyBased}

\subsection{Background on Time-expanded Network}

We start introducing \framework\ by bringing the notion of Time-expanded Network (TEN) to the distributed ML domain.  
TEN is widely used in flow optimization problems~\cite{tenUse1, tenUse2, tenUse3} because it enables judicious consideration of both network topology and timing information in a single data structure.

A visual example of a homogeneous, asymmetric 3-NPU topology is shown in~\autoref{fig:HeteroTenDefinition}(a), as well as its \ten\ representation in~\autoref{fig:HeteroTenDefinition}(b).  
Each NPU in the base topology forms a column of vertices in the \ten\ representation, with each column corresponding to a unique time span.  
This column is then replicated across a time range (up to $t$=3 in this example).  
The topology exhibits four unidirectional connections: 1$\rightarrow$2, 1$\rightarrow$3, 2$\rightarrow$3, and 3$\rightarrow$1.  
These connections translate into distinct edges in the \ten\ representation, linking different time spans flowing from left to right.  

This \ten\ representation effectively integrates the temporal and spatial information of the network communication behavior into a single data structure.  
Consider the scenario where a chunk starts being transmitted from NPU 1 to NPU 2 at $t$=1.  
It can be seen from the \ten\ that this chunk will reach its destination at $t$=2.  
Similarly, the absence of a direct link from NPU 3 to NPU 2 in the \ten\ illustrates that such a direct transmission is impossible due to the lack of the physical link in the network.

\insertFigure{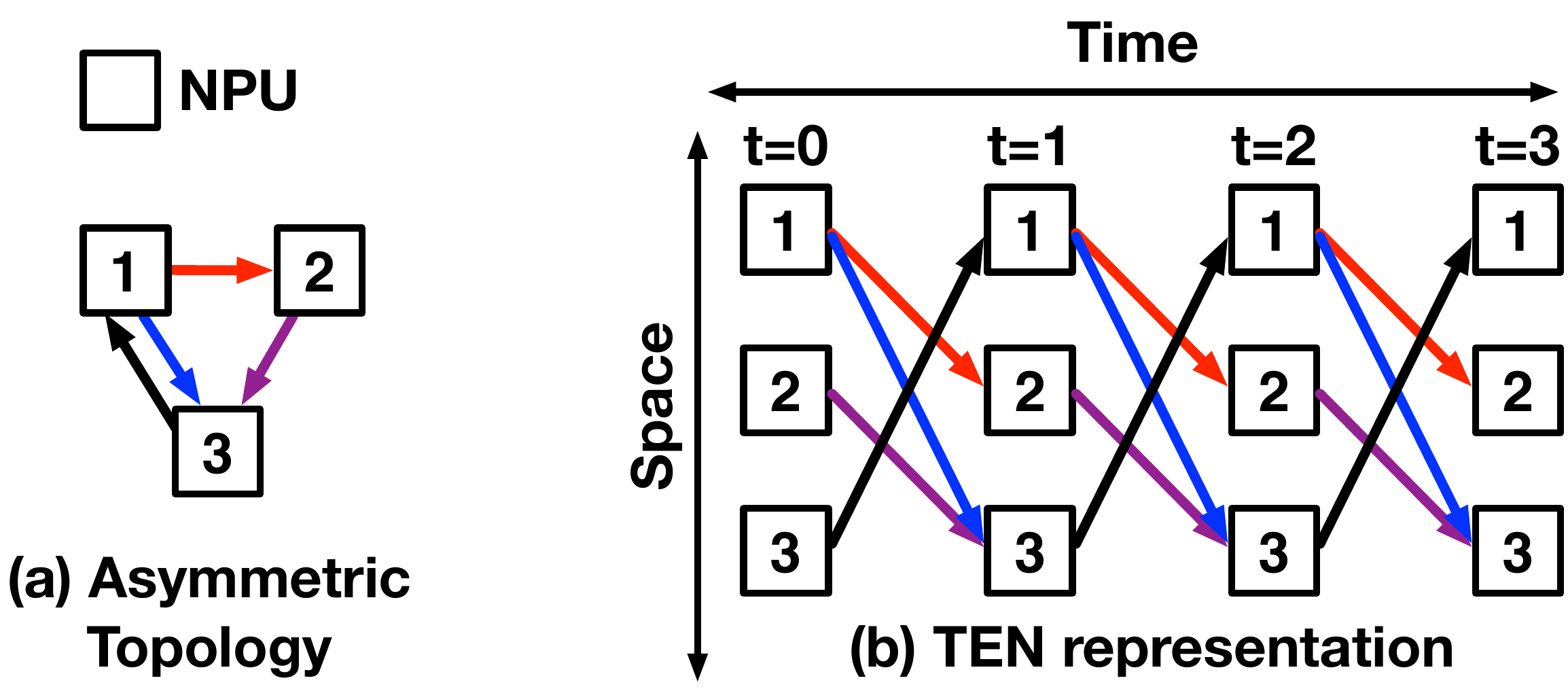}{  
(a)~An homogeneous, asymmetric topology with 3 NPUs and 4 links.  
(b)~\ten\ expansion of the network up to $t$=3.
}{0.85}{-1em}{0em}

\insertFigure{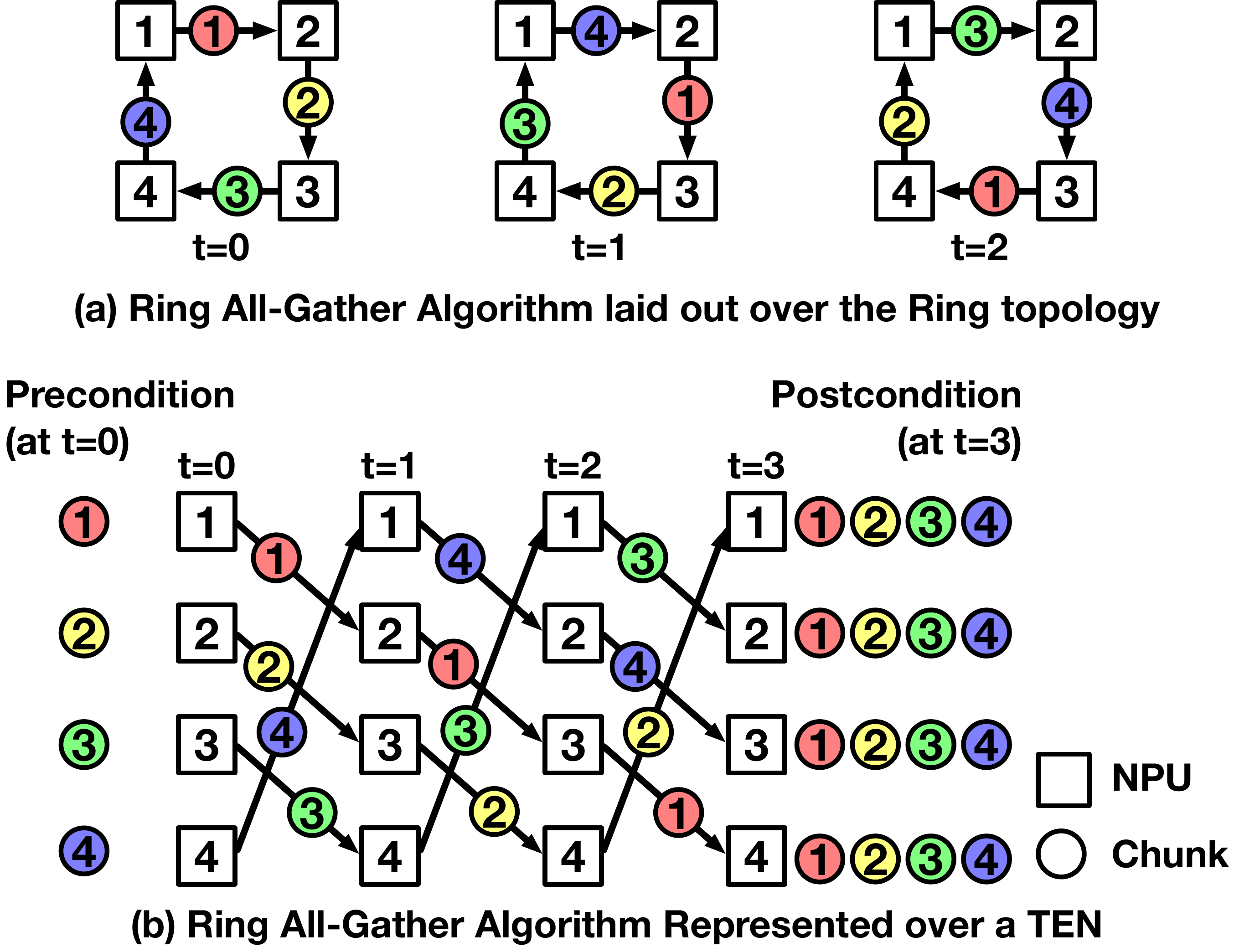}{
(a)~Unidirectional \ringAlg\ \allgather\ algorithm laid out over a unidirectional \ring\ topology across different time spans.
(b)~Algorithm shown in~(a) captured using the \ten-based representation.
}{0.93}{-0.8em}{-1em}

\subsection{Representing Collective Algorithms using TEN}

\ten\ offers a concise mechanism to represent collective algorithms in a single data structure.  
One example is shown in~\autoref{fig:CollectiveOnTEN}.  
Depicted in~\autoref{fig:CollectiveOnTEN}(a) is an unidirectional \ringAlg\ \allgather\ algorithm laid over the \ring\ topology across distinct time spans.  
The same collective algorithm is represented over the \ten\ representation in~\autoref{fig:CollectiveOnTEN}(b).  
The leftmost segment of the \ten\ depicts the \emph{precondition} (i.e., chunks initially held by each NPU) at $t$=0, while the rightmost segment represents the \emph{postcondition} (i.e., chunks held at the end) at $t$=3.

In this visualization, a chunk transmission between two NPUs is symbolized by the chunk occupying a \ten\ link, which we term a \emph{link-chunk match}.  
For this example, all links in the \ten\ structure correspond to matched chunks, indicating maximized resource utilization throughout the course of the collective execution.  
Furthermore, \emph{each link matches up to one chunk}, eliminating network contention from multiple payloads assigned over a single link at a time.

To the best of our knowledge, this work is the first to introduce \ten\ to the domain of distributed ML and to represent collective algorithms atop it.  
We believe this formulation has value beyond the specific collective synthesis mechanism presented in this paper.

\insertFigure{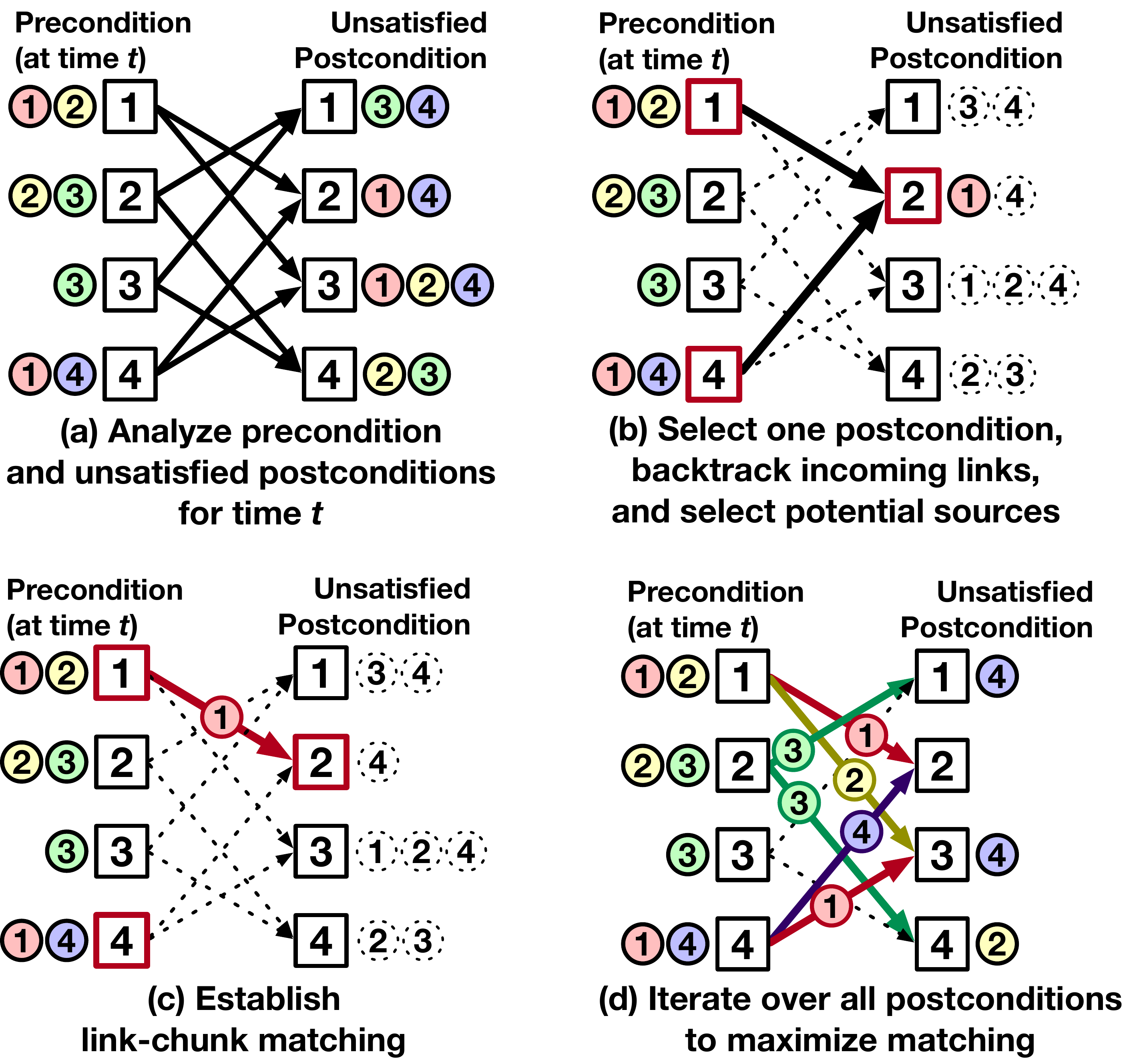}{
Network Utilization Maximizing Matching algorithm.
For a given time span $t$ of a \ten:
(a)~The pre/postconditions of a collective communication are evaluated.
(b)~An unsatisfied postcondition is randomly selected, and the destination NPU is backtracked over the \ten\ to identify candidate source NPUs capable of providing the requested chunk.
(c)~One candidate is randomly chosen and a link-chunk match is made.
(d)~Steps (b)-(c) are repeated exhaustively over all unsatisfied postconditions to maximize the number of link-chunk matches, optimizing link utilization for the given time $t$.
}{1}{-2em}{-1em}

\subsection{TACOS' Approach to Collective Algorithm Synthesis}

The \ten-based representation employed by \framework\ enables it to frame the collective synthesis problem as a link-chunk match-making challenge, rather than as a global optimization problem.  
This strategy offers a significant advantage, as solving global optimizations (e.g., Integer Linear Programming (ILP) formulations) is an NP-hard problem~\cite{ilpNPHard}.

Instead, in order to synthesize a collective algorithm, \framework\ aims to \emph{establish and maximize the number of link-chunk matches} for individual chunks onto the \ten\ links.  
To synthesize the collective algorithm, \framework\ first constructs a \ten\ for $t$=0 for the target network topology.  
It then proceeds to \emph{execute a link-chunk matching algorithm} (described in~\autoref{sec:matching-algo}) to maximize network utilization at $t$=0.  
The \emph{\ten\ is then expanded by one more time span}, and the process repeats until all postconditions are met.

This iterative synthesis approach maximizes network utilization over the entire course of the collective execution, thereby producing a high-performance collective algorithm.  
Moreover, the framework \emph{inherently mitigates link congestion} by allowing each \ten\ link to accommodate only one chunk.  
Importantly, \framework\ supports arbitrary networks since any heterogeneous or asymmetric topologies can still be represented in \ten, eliminating the need for specific homogeneity or symmetry assumptions.

\insertFigure{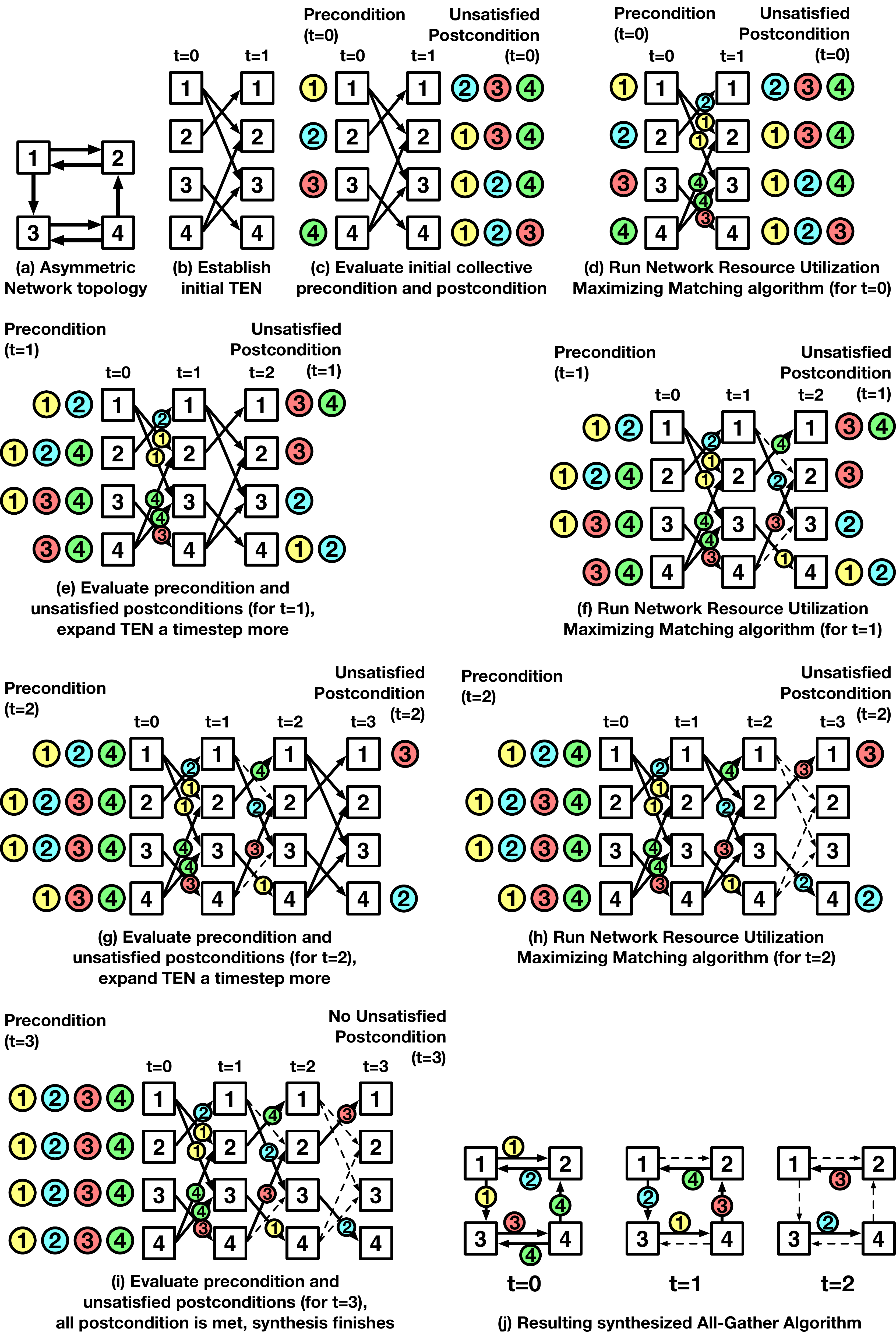}{
End-to-end overview of how \framework\ constructs an \allgather\ algorithm for a homogeneous, asymmetric 4-NPU topology.
}{1}{-2em}{-1em}

\subsection{Network Utilization Maximizing Matching Algorithm}\label{sec:matching-algo}

\framework\ employs a swift \emph{network utilization maximizing matching algorithm} to find link-chunk matches for a given time span $t$.  
This algorithm is summarized in~\autoref{fig:GreedyMatching}.  
The matching process begins in~\autoref{fig:GreedyMatching}(a), where \framework\ evaluates the preconditions and unsatisfied postconditions of each NPU at the given time span $t$ to determine the chunks requested to be received by each NPU.  
\framework\ then randomly selects one unsatisfied postcondition to evaluate potential link-chunk matches.  
For instance, in~\autoref{fig:GreedyMatching}(b), NPU 2 and chunk 1 are selected.  
The algorithm then backtracks the \ten\ graph to identify a set of potential source NPUs capable of supplying the requested chunk, as represented by the bold arrows in the figure.  
These source NPUs are assessed to determine which NPUs currently possess and can provide the desired chunk.
In the illustrated scenario, both NPU 1 and NPU 4 fulfill these criteria.  
From this pool of candidates, one source NPU is randomly selected, resulting in a successful link-chunk match as depicted in~\autoref{fig:GreedyMatching}(c).  
In this example, NPU 1 is chosen to cater to chunk 1, and the corresponding \ten\ link (NPU 1 to 2) becomes occupied by the established match.  
This sequence repeats for all unsatisfied postconditions, yielding the outcome in~\autoref{fig:GreedyMatching}(d).  
The algorithm is summarized in~\autoref{alg:ten_matching}.  

For a given time span $t$, the matching algorithm exhaustively iterates over all unsatisfied postconditions and makes random selections whenever possible.
Therefore, the algorithm tries to \textit{maximize the number of successful link-chunk matches}, thereby ensuring nearly maximal utilization of network link resources at the given time $t$.  
Additionally, as only one chunk can be matched over a link, this algorithm effectively eliminates link congestion.

\input{algorithm/ten_matching.tex}

\input{algorithm/ten_e2e.tex}

\insertFigureWide{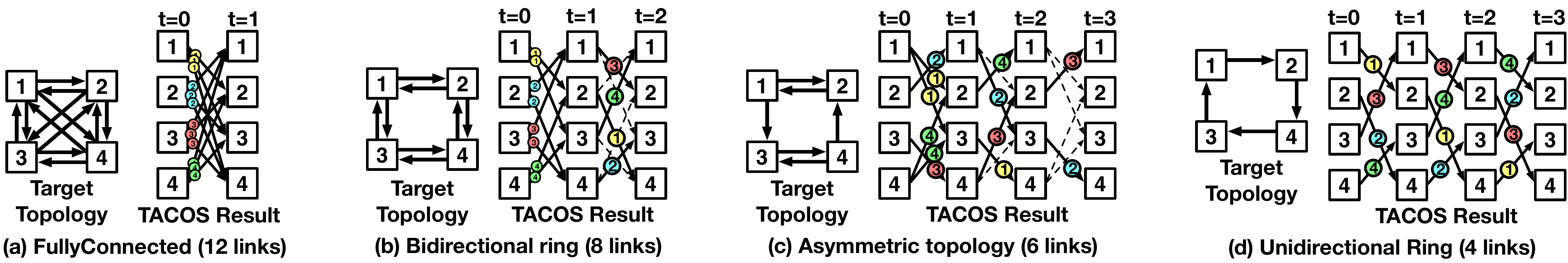}{
Distinct target topologies with 4 NPUs but different numbers of links, and their \allgather\ synthesis results over \ten\ using \framework.  
As the network connection becomes more sparse, \framework\ is required to expand the \ten\ for more time spans to satisfy all collective postconditions.  
Still, note that the matching algorithm was able to maximize the network utilization for every time span.
}{1}{-2em}{-1em}

\subsection{End-to-end \framework\ Collective Synthesis}

\insertFigure{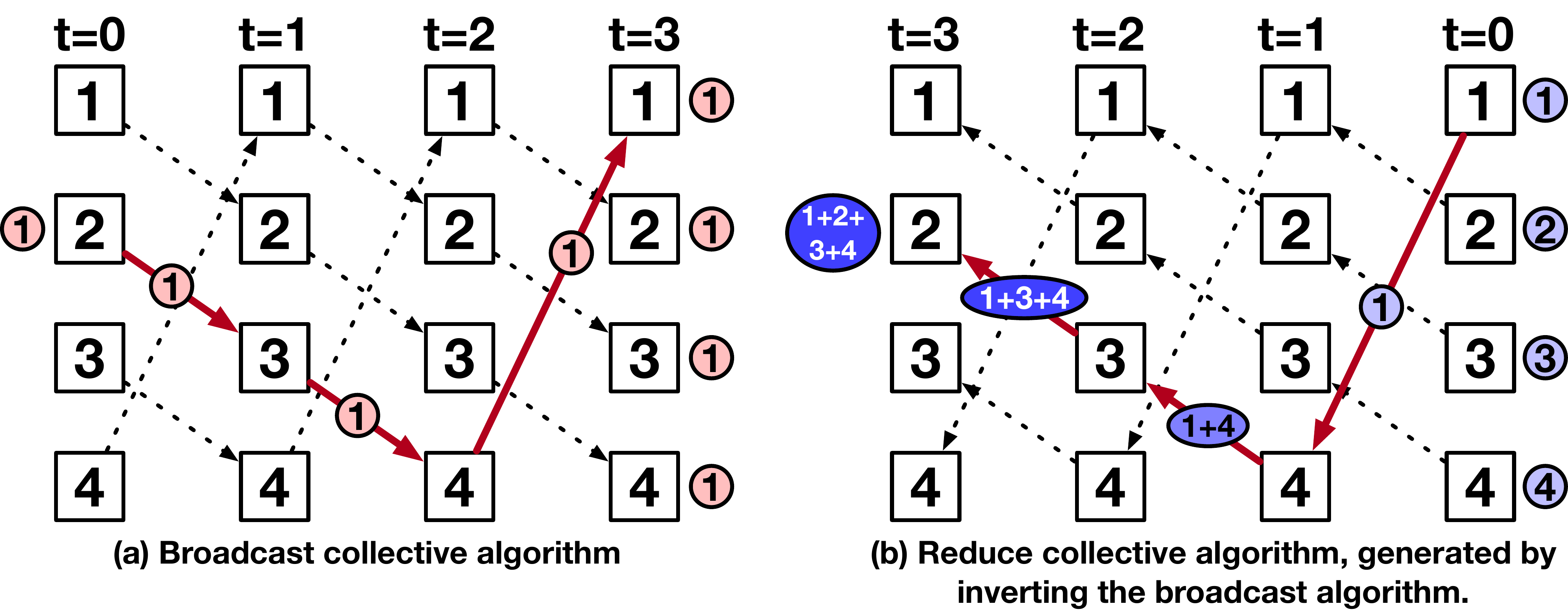}{
Synthesis of collectives with reduction operations (e.g., \reducescatter).
(a)~\framework\ first constructs a TEN whose $(src, dest)$ are reversed to $(dest, src)$ (reversed link directions) and synthesizes corresponding non-reduction collectives.
(b)~Then, \framework\ reverses the synthesized collectives back to get the final combining collectives.
}{1}{-2em}{-1em}

To help understand the end-to-end synthesis process, \autoref{fig:TACOSInDepth} provides an illustrative example synthesizing an \allgather\ algorithm for a 4-NPU asymmetric network.  
\autoref{fig:TACOSInDepth}(a) depicts the target topology, while \autoref{fig:TACOSInDepth}(b) shows the initial \ten\ representation of the network for $t$=0.  
First, preconditions and postconditions are evaluated for $t$=0, as shown in~\autoref{fig:TACOSInDepth}(c).  
Subsequently, the link-chunk matching algorithm is executed, resulting in maximized network utilization for $t$=0 as demonstrated in~\autoref{fig:TACOSInDepth}(d).  
The preconditions and postconditions are updated for $t$=1, and since there are remaining unsatisfied postconditions, the \ten\ is expanded for one more time span, resulting in the structure depicted in~\autoref{fig:TACOSInDepth}(e).  
This process repeats iteratively: the link-chunk matching algorithm is executed for $t$=1, once again maximizing network utilization, as shown in~\autoref{fig:TACOSInDepth}(f).  
The pre/postcondition re-evaluation, \ten\ expansion, and link-chunk matching are repeated for $t$=2, as depicted in~\autoref{fig:TACOSInDepth}(g--i).  
Once all postconditions are satisfied, the synthesis terminates at \autoref{fig:TACOSInDepth}(i).  
The resultant collective algorithm, laid out on the original network topology, is summarized in~\autoref{fig:TACOSInDepth}(j), ready to be deployed by CCLs.  

This end-to-end synthesis algorithm is summarized in~\autoref{alg:ten_e2e}.  
We first initialize TEN and postconditions for $t=0$.  
Until all postconditions are met, \framework\ simply expands the TEN by one more time span and runs the utilization maximization matching process.  
Since \framework\ only schedules communications between neighboring NPUs, the \framework-synthesized algorithm is deadlock-free~\cite{multitree}.  

\autoref{fig:TacosTopologyComparison} illustrates how \framework\ synthesis results react to changes in the underlying topology.  
All target topologies in~\autoref{fig:TacosTopologyComparison}(a--d) have 4 NPUs but different numbers of physical links connecting them.  
For the \fullyconnected\ topology, all postconditions can be satisfied in a single shot, effectively yielding a \direct\ algorithm.  
As connectivity becomes scarcer, the execution of the collective takes longer.  
It is important to note, however, that \framework\ remains effective in maximizing link utilization for each discrete time span, resulting in optimal collective algorithms with minimized collective time.

\noindent \textbf{Collectives with Reduction Operations.}
Certain collectives, such as \reducescatter\ or \allreduce, involve reduction operations among chunks.
\framework\ handles such collectives by synthesizing their non-combining counterpart collective patterns.
An illustrative example with Broadcast and Reduce collectives is shown in~\autoref{fig:CombiningCollective}. 
To synthesize Reduce collective, all $(src, dest)$ links are inversed into $(dest, src)$ and the corresponding Broadcast algorithm gets synthesized.
By reversing the synthesized algorithm back, \framework\ can obtain the Reduce algorithm.
Similarly, search for a \reducescatter\ algorithm can be accomplished by synthesizing an \allgather\ algorithm and reversing the sender-receiver of all communication pairs.
An \allreduce\ operation can be synthesized by combining a \reducescatter\ followed by an \allgather.

\insertFigure{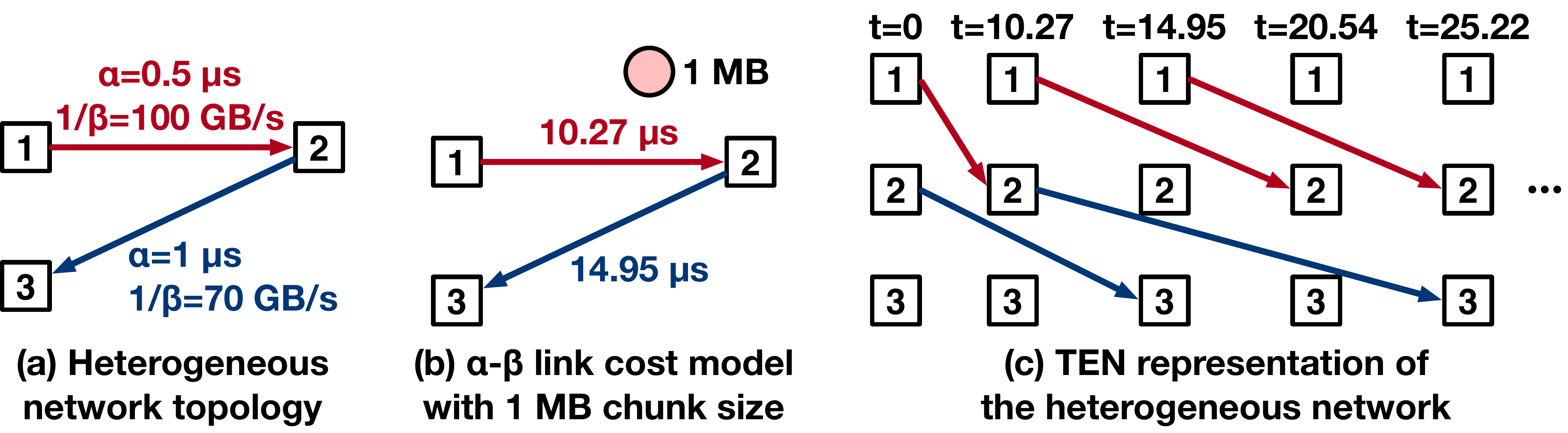}{
(a)~A heterogeneous 3-NPU topology. (b)~$\alpha$-$\beta$ model applied to calculate link costs. (c)~\ten\ representation of the heterogeneous topology, expanded up to $t$=25.22~µs.
}{1}{-2em}{-1em}

\subsection{\framework\ for Heterogeneous Networks}

The basics of conducting link-chunk matching over \ten\ remain consistent even for heterogeneous networks.

\noindent \textbf{Heterogeneous Topology in \ten.}
To delineate heterogeneous link costs, we adopt the $\alpha$--$\beta$ model~\cite{alphabeta, libra, themis, sccl, taccl}.
$\alpha$ characterizes the latency of the link, whereas $\beta$ represents the reciprocal of link bandwidth (i.e., serialization delay per unit message size).
Given a chunk size $n$, we can translate each link transmission cost into a single number, $\alpha + \beta \times n$.
Once the link cost is set, a \ten\ representation can be drawn.
\autoref{fig:NetworkRepresentation} shows this process.
The network configuration in~\autoref{fig:NetworkRepresentation}(a) features a heterogeneous topology.
Each link delay is derived using the $\alpha$--$\beta$ model in~\autoref{fig:NetworkRepresentation}(b) (assuming a 1~MB chunk in this scenario).
\autoref{fig:NetworkRepresentation}(c) shows the resultant \ten\ representation, which can then be absorbed by the \framework\ framework.
Although we have not considered NPU or memory performance during synthesis in this work, these factors can be also incorporated into the $\alpha$ and $\beta$ costs of each link.

\noindent \textbf{Prioritizing Lower-cost Links.}
When conducting link-chunk matching, multiple links with different transmission costs may emerge as potential matches due to network heterogeneity.
To minimize collective time, \framework\ prioritizes the link with the lowest cost when making link-chunk matches.

\subsection{\framework\ for \switch-based Networks}\label{subsec:tacos_extension}

\insertFigure{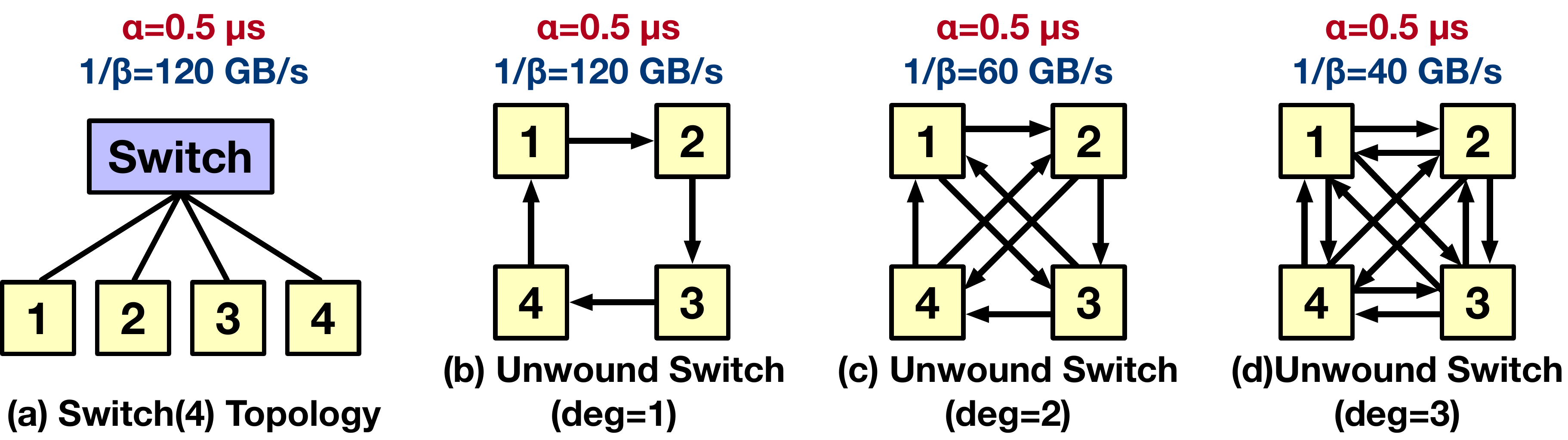}{
(a)~A \switch\ fabric with 4 NPUs. (b)-(d)~Unwinding the switch network with degrees 1 through 3.
}{1}{-2em}{-1em}

A \switch\ topology offers versatile NPU connectivity, but unregulated use leads to sub-optimal performance due to network contentions or even deadlocks.
To address this, we propose a mechanism to unfold \switch\ networks into fixed, point-to-point connections.
With a degree $d$ unwinding approach for an $N$-NPU \switch, we establish $d$ outgoing links from each NPU $i$, connecting to NPUs ($i$+1), ($i$+2), $\cdots$, ($i$+$d$) (mod $N$).\footnote{
This unwinding scheme neither covers all potential point-to-point configurations nor dynamic \switch\ connectivity across \ten\ time spans.
Notably, $d$=1 and $d$=$N$-1 unwinding are suitable for bandwidth and latency-critical collective synthesis, respectively~\cite{taccl}.
However, developing a more flexible \switch\ unwinding methodology remains a subject for future investigation.
}
During this transition, the cost $\alpha$ associated with each point-to-point link remains consistent.
However, due to shared link bandwidth, the $\beta$ cost multiplies by $d\times$.
\autoref{fig:UnwindSwitch}(a) illustrates a 4-NPU switch while \autoref{fig:UnwindSwitch}(b--d) depict its corresponding $d$=1--3 unwinding.
Once the switch is unwound, the topology can be represented in \ten\ and used by \framework.

%% file: algorithm/ten_matching.tex
\begin{figure}[t]
\vspace{-0.7em}
\begin{algorithm}[H]
\caption{Utilization Maximizing Matching Algorithm}
\label{alg:ten_matching}
\begin{algorithmic}

\State \textbf{Condition:} Provided TEN at time $t$
\State

\State // Get unsatisfied postconds
\For{$d$ $\leftarrow$ $\forall$NPUs}
\For{$c$ $\leftarrow$ $\forall$Chunks}

\If{postcond[$d$][$c$] \&\& !precond[$d$][$c$]} 
    \State Add ($d$, $c$) to unsatisfied\_postcond 
\EndIf

\EndFor
\EndFor

\State
\State // Make link-chunk matches
\State random\_shuffle(unsatisfied\_postcond)
\For{($d$, $c$) $\leftarrow$ $\forall$unsatisfied\_postcond}

\State // Select $s$ NPU that can provide $c$ to $d$
\State s $\leftarrow$ backtrack( TEN[$t$][*][$d$] )
\State candidate\_srcs $\leftarrow$ List($\forall s$: precond[$s$][$c$] is True)
\State chosen\_src $\leftarrow$ random\_select(candidate\_srcs)

\State
\State // Mark link-chunk match
\State precond[$d$][$c$] $\leftarrow$ mark True
\State TEN[$t$][chosen\_src][$d$] $\leftarrow$ mark $c$

\EndFor

\end{algorithmic}
\end{algorithm}
\vspace{-1em}
\end{figure}

%% file: algorithm/ten_e2e.tex
\begin{figure}[t]
\vspace{-0.7em}
\begin{algorithm}[H]
\caption{\framework\ End-to-End Synthesis}
\label{alg:ten_e2e}
\begin{algorithmic}

\State \textbf{Condition:} Provided target topology and collective
\State \textbf{Initialize:} TEN[$t=0$], pre/postconds
\State

\While{$\exists$unsatisfied\_postconds}

\State // Expand TEN by 1 time span and run matching
\State $t \leftarrow t+1$
\State Expand TEN[$t$]
\State Utilization\_Maximize\_Matching(TEN[$t$]) (\autoref{alg:ten_matching})

\EndWhile

\end{algorithmic}
\end{algorithm}
\vspace{-2em}
\end{figure}

%% file: content/5_methodology.tex
\section{Methodology}\label{sec:methodology}

\subsection{Baseline Collective Algorithms and Synthesizers}

For performance evaluations, we compare \framework\ against the following baseline \allreduce\ algorithms.

\begin{itemize}[leftmargin=*]
    \item \textbf{\ringAlg, \direct.}
    These are basic collective algorithms showing traffic patterns depicted in \autoref{fig:CollectiveAlgorithmPattern}.
    In particular, \ringAlg\ currently serves as the default algorithm across most CCLs~\cite{nccl, oneCCLDoc, rccl}.
    
    \item \textbf{\hd, \dbt.}
    We also assessed \hd~\cite{rabenseifner} and \dbt~\cite{dbt} in \autoref{table:smallScaleResult} since they are only suited for networks with a power-of-two number of NPUs.
    
    \item \textbf{\hierarchical, \themis, \ccube.}
    These are topology-aware collective algorithms manually crafted to accommodate specific network topologies.
    \hierarchical~\cite{blueconnect} develops a multi-rail \allreduce\ algorithm for a symmetric hierarchical network.
    \themis~\cite{themis} further optimizes \hierarchical\ through improved chunk-level scheduling. 
    Meanwhile, \ccube~\cite{ccube} manually maps two binary trees on DGX-1 and executes two tree-based \allreduce\ algorithms in parallel.
    
    \item \textbf{\multitree, \taccl.}
    \multitree~\cite{multitree} synthesizes collective algorithms for homogeneous networks via spanning tree construction, while \taccl~\cite{taccl} employs an ILP-based approach for symmetric heterogeneous topologies.\footnote{
    \ccube\ and \multitree\ implementations are not public, and we used their reported numbers in the paper for an apples-to-apples comparison. Since \taccl\ offers limited topology options, we implemented a \taccl-like baseline by integrating its ILP formulation over our \ten\ representation.
    }
    
    \item \textbf{Ideal.}
    Finally, to show \framework' absolute synthesis quality, we also augmented all results with theoretically ideal collective performance.
    This can be derived from the topology diameter (the minimum latency for the farthest two NPUs to communicate) for $\alpha$ costs, and the bottleneck serialization delay (injection and ejection time of all chunks) for $\beta$.

    \begin{equation*}
        \text{Ideal} = \frac{\text{CollectiveSize} \times 2(n-1)/n}{\min_{N \in \forall \text{NPU}} (\text{BW}_N)} + \text{Diameter}
    \end{equation*}
\end{itemize}

\input{table/target_topology.tex}

\subsection{Target Topologies and Synthesis Environment}

\autoref{table:targetTopology} lists the topologies studied in this work, covering both homogeneous and heterogeneous networks, as well as both symmetric and asymmetric topologies.\footnote{
Unless otherwise specified, we set link $\alpha$=0.5 µs and 1/$\beta$=50 GB/s.
}
\framework\ synthesis time is measured using the Intel Xeon E5-2699v3 CPU.

\subsection{Simulation Infrastructure}

For this work, a simulator was necessary to study performance across the diverse spectrum of topologies that we examined (\autoref{table:targetTopology}).
We use the ASTRA-sim distributed ML simulator~\cite{astrasim, astrasim2} to evaluate the performance of the baseline collective algorithms and their corresponding impact on full workloads.
ASTRA-sim includes NCCL-validated implementations of the basic collective algorithms.
It also models chunking (i.e., breaking a large collective into smaller chunks) and its related effects in \hierarchical\ and \themis.

\noindent \textbf{Network Simulation Backend.}
To model the network oversubscription and congestion behaviors incurred by topology-unaware collectives, while enabling the simulation of large-scale network topologies, we enhanced the network modeling of ASTRA-sim by creating and attaching a congestion-aware analytical network simulation backend.
The congestion-aware analytical backend simulates a message transfer by simulating the send and receive operations at the link granularity.
Each link is equipped with message queues and can process (i.e., send and receive) only one message at a time.
For example, if two messages contend for the same link, only one message is sent out in a first-come, first-served order.
Such simulation infrastructure enables the first-order modeling of network queuing and congestion effects for networks at scale.

\noindent \textbf{Validation.}
We validated ASTRA-sim's baseline \allreduce\ implementations over two real systems: an NVLink-based, 8-GPU V100 server and a 32-TPUv3 (8$\times$4) cluster.
We observed ASTRA-sim to be 6--8\% off on average, with the difference close to 1\% for large message sizes~\cite{astrasimvalidation}.

%% file: table/target_topology.tex
\begin{scriptsize}
\begin{table}[t]
\centering
\caption{
Topologies evaluated in this paper.
}
\label{table:targetTopology}
\vspace{-0.5em}
\begin{tabular}{|c|c|c|}
\hline
\textbf{Topology} & \textbf{Heterogenous} & \textbf{Asymmetric} \\ \hline
Ring (RI) &  & \\ \hline
FullyConnected (FC) &  & \\ \hline
2D Torus &  & \\ \hline
3D Torus &  & \\ \hline
2D Mesh (Mesh) &  & $\checkmark$ \\ \hline
3D Hypercube (HC) &  & $\checkmark$ \\ \hline
2D Switch & $\checkmark$ &  \\ \hline
3D Ring-FC-Switch (\rfs) & $\checkmark$ &  \\ \hline
\dragonfly\ (\df) & $\checkmark$ & $\checkmark$ \\ \hline
\end{tabular}
\vspace{-1.5em}
\end{table}
\end{scriptsize}

%% file: content/6_result.tex
\section{Evaluations}

\subsection{TACOS Synthesis Result}

\insertFigure{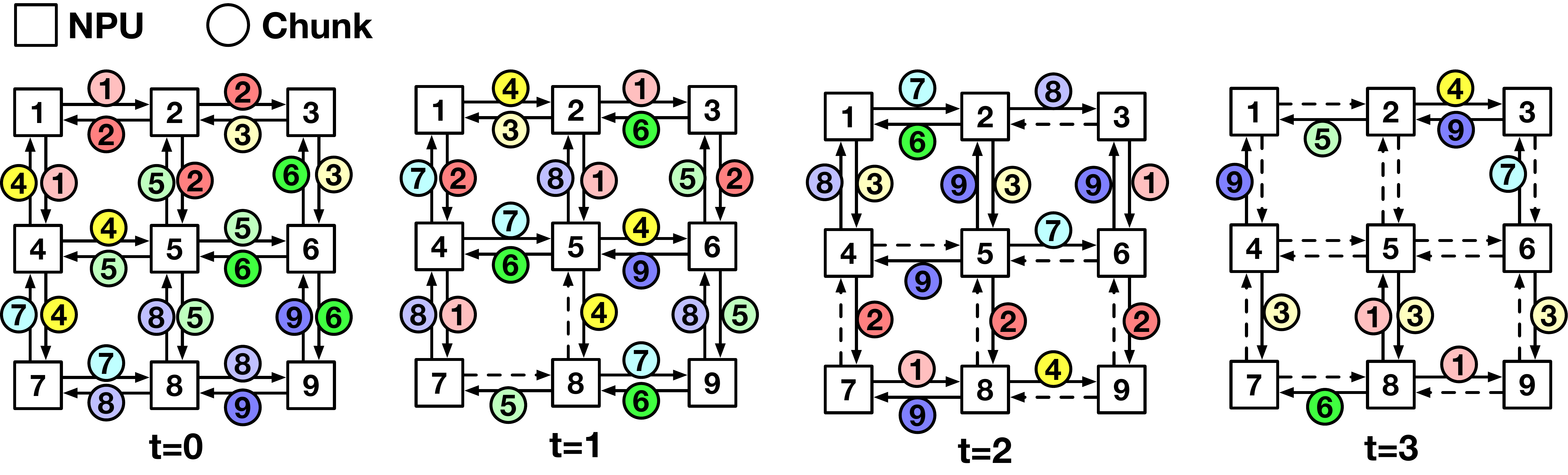}{
An example \allgather\ collective algorithm synthesized by \framework\ over a homogeneous $3\times3$ \mesh\ topology.
}{1}{-2em}{-1em}

We begin our evaluation with a visual representation of a \framework-synthesized algorithm. Specifically, we synthesized an \allgather\ algorithm for a $3\times3$ \mesh\ network, as shown in \autoref{fig:MeshExampleSearchResult}. The synthesized algorithm effectively avoids network contentions, demonstrating \framework' ability to autonomously alleviate link congestion during synthesis.

\subsection{Analysis of TACOS-Synthesized Algorithms}

Modern distributed ML clusters typically employ symmetric topologies, whether homogeneous~\cite{google2023tpupod,cloudTpuPaper,tpuv4} or heterogeneous~\cite{themis, NVIDIASuperPod, HabanaPtP}.
This section highlights the benefits of \framework\ in handling these regular, symmetric topologies, compared to baseline collective algorithms and existing solutions.

\insertFigure{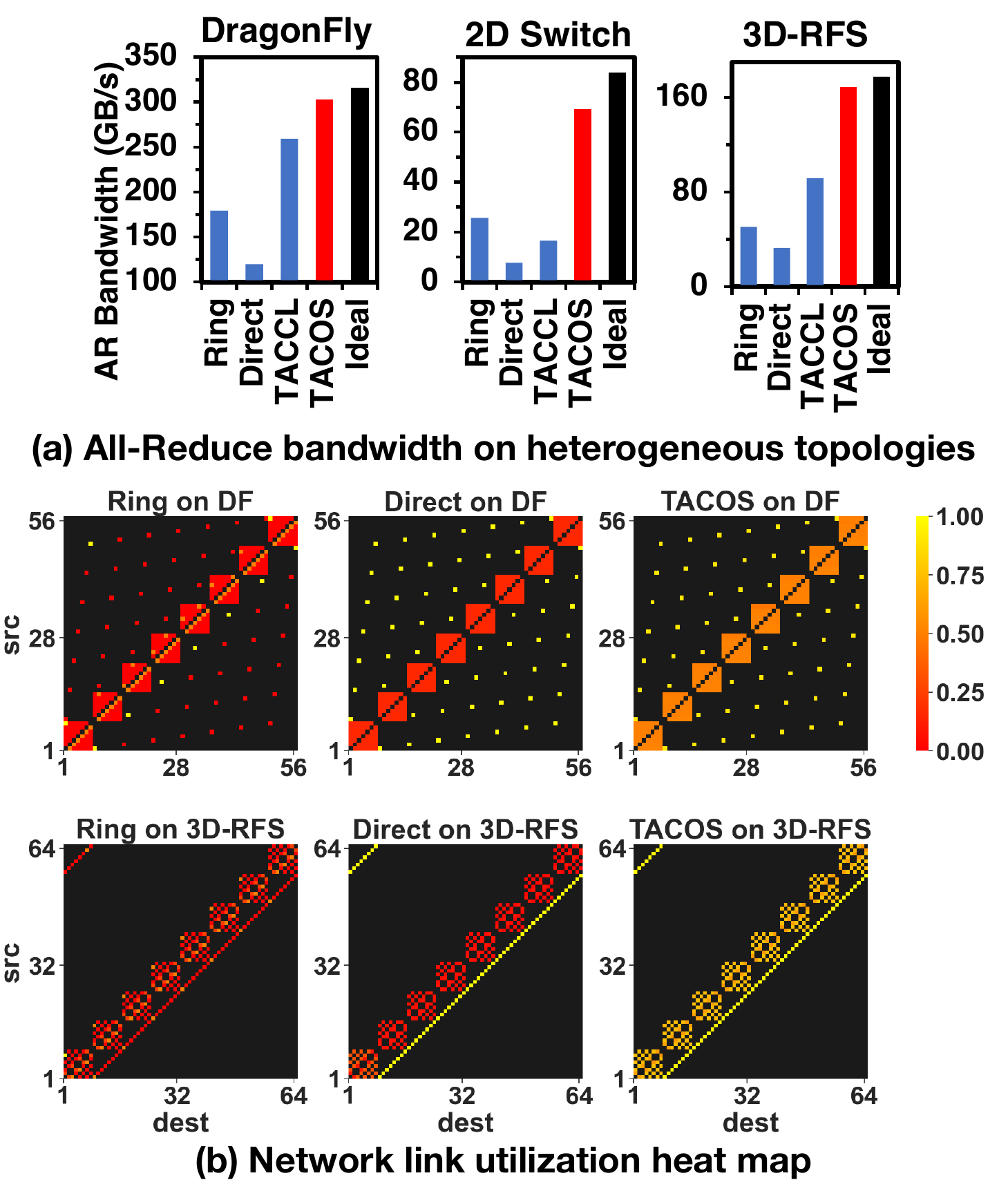}{
(a)~\allreduce\ bandwidth of \ringAlg\ and \direct\ basic algorithms, and \taccl\ and \framework-synthesized algorithms.
(b)~Average link utilization measured on \dragonfly\ and \rfs.
\framework\ achieved 90.84\% efficiency compared to the theoretical ideal.
}{1}{-2em}{-0.5em}

\noindent \textbf{1. Topology Exploration.}
We synthesized topology-aware \allreduce\ algorithms for three distinct systems: a \dragonfly\ topology ($4\times5$) with link bandwidths of [400, 200] GB/s (per dimension), a 2D \switch\ ($8\times4$) with [300, 25] GB/s, and \rfs\ (2$\times$4$\times$8) with [200, 100, 50] GB/s.
Notably, the \dragonfly\ topology is both asymmetric and heterogeneous, while \switch\ and \rfs\ are symmetric topologies.

\autoref{fig:TopologyExp}(a) illustrates the \allreduce\ bandwidth (i.e., collective size $\div$ collective time) synthesized by \framework, alongside \ring, \direct, and \taccl-synthesized algorithms.
\framework\ effectively synthesizes collective algorithms, resulting in an average speedup of $2.56\times$ over the baseline algorithms, including symmetric topologies.
Moreover, thanks to its congestion-free search mechanism, \framework\ even outperforms \taccl, consistently achieving more than 90\% efficiency compared to the theoretical upper bound.

The heat map in \autoref{fig:TopologyExp}(b) shows the average link utilization across all links for the \dragonfly\ and \rfs\ networks. Topology-unaware basic algorithms lead to significant oversubscription on certain links, leaving others underutilized. However, the topology-aware collective algorithms synthesized by \framework\ improve overall link utilization, distributing traffic more evenly across all links.

\input{table/small_scale_result.tex}

\insertFigure{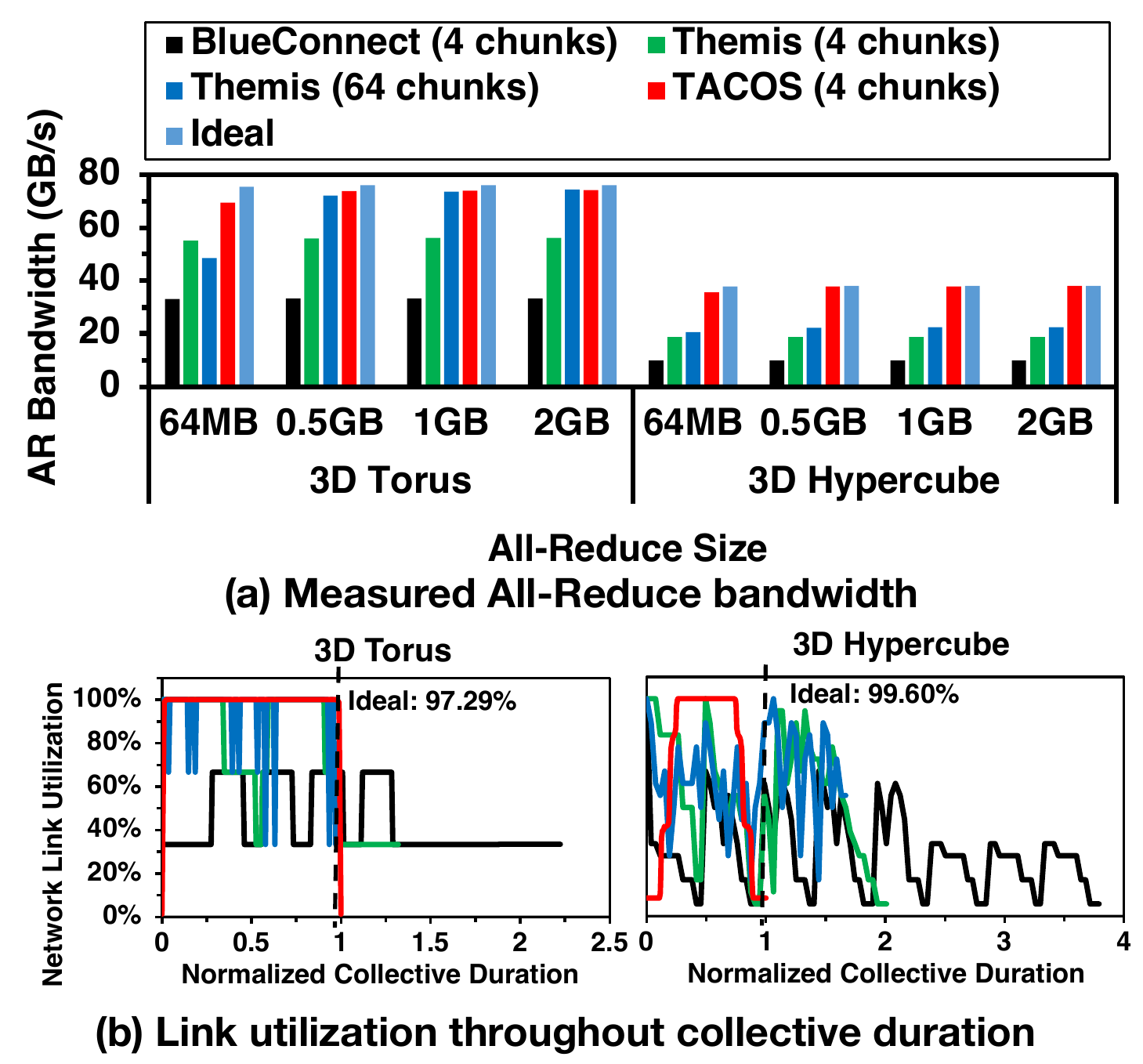}{
(a)~\allreduce\ bandwidth comparison of \framework\ against \hierarchical\ and \themis. 
\framework\ achieves an average of 97.00\% efficiency relative to the theoretical bound, regardless of collective size, while optimizations like \themis\ may struggle with latency-critical collectives. 
(b)~Link utilization rate during collective execution, normalized by \framework\ collective time. 
\framework\ achieves 98.44\% efficiency on average, regardless of topology shape.
}{1}{-2em}{-1em}

\noindent \textbf{2. Multi-node Analysis.}
To demonstrate \framework' applicability for multi-node, symmetric AI clusters, we focus on the \rfs\ topology and scale it by increasing the size of the last dimension (i.e., node). 
For instance, a 16-node system has 2$\times$4$\times$16=128 NPUs. 
\autoref{table:smallScaleResult} provides a comprehensive summary of these evaluations.
For a 128-NPU cluster, \taccl\ incurred intractable synthesis times due to its NP-hard approach.
On average, the collective algorithm generated by \framework\ exhibited a 5.39$\times$ speedup over the \ringAlg\ baselines across all configurations.
\direct\ baselines suffered as the \direct\ algorithm caused heavy network contention across the topology.
\framework\ achieved 75.88\% efficiency relative to the theoretical ideal.

\noindent \textbf{3. BlueConnect and Themis.}  
To evaluate \framework' performance against manually designed topology-aware collectives, we compare its results with the \hierarchical~\cite{libra} and \themis~\cite{themis} algorithms.  
The \hierarchical\ algorithm is designed for executing \allreduce\ on symmetric multi-dimensional networks by applying \reducescatter\ and \allgather\ sequentially across each network dimension.
\themis\ improves \hierarchical's efficiency by allowing distinct chunks to traverse network dimensions in an arbitrary manner, thereby balancing the load.  
We assessed \framework' performance over a \torus\ topology with $\alpha$=0.7 µs and 1/$\beta$=25 GB/s links.

We measured \framework' benefits over \themis' target topology, a symmetric 3D Torus.  
The \allreduce\ bandwidth is summarized in~\autoref{fig:ResultThemisRuntime}(a).  
\framework\ achieved 95.90\% efficiency relative to the ideal case.  
\themis, using 64 chunks, achieved similar efficiency for large collectives, but this came at the cost of latency.  
For latency-critical small collectives, \themis' efficiency dropped to 64.37\%.  
When only 4 chunks per collective were used to avoid this issue, \framework\ consistently outperformed \themis.  
Furthermore, for \themis-unfriendly asymmetric topologies like the 3D Hypercube, \themis\ achieved only 49.43\% efficiency, while \framework\ reached 98.10\% of the ideal case.  
This discrepancy arises because \themis\ cannot alter the path each chunk takes, limiting its ability to mitigate network contention when the optimal collective path for each network dimension is unknown.  
Consequently, \framework\ demonstrated an average \allreduce\ bandwidth 2.01$\times$ higher than \themis.  
\autoref{fig:ResultThemisRuntime}(b) illustrates the link utilization during the collective execution.  
For the symmetric \torus, both \framework\ and \themis\ achieved nearly 100\% utilization.  
However, in the \hypercube\ topology, \themis\ experienced significant fluctuations in utilization due to network contention, which \framework\ successfully avoided.  
We emphasize that \framework\ achieved these results autonomously, without requiring any manual design efforts.

\insertFigure{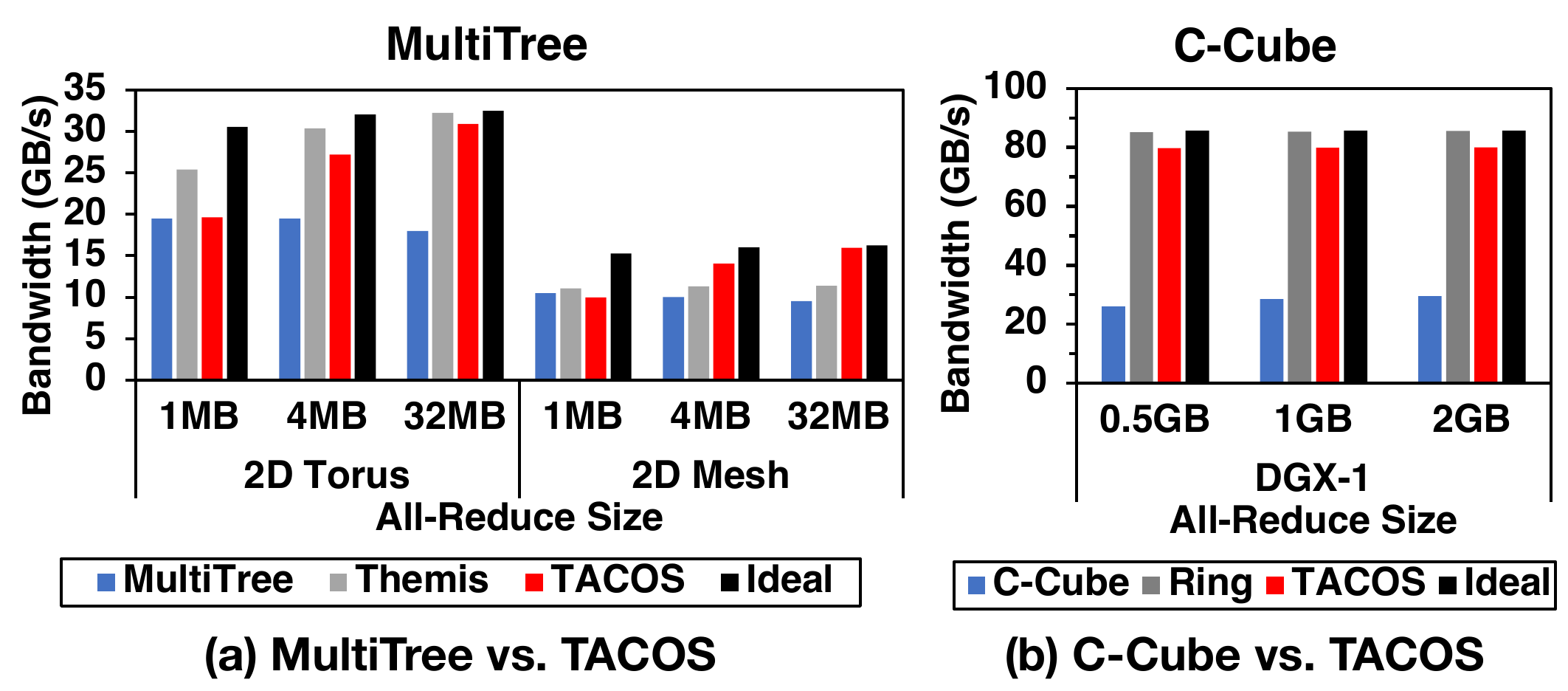}{
\allreduce\ bandwidth of \framework\ and relative speedups compared to (a)~\multitree-synthesized and (b)~\ccube\ algorithms.  
\framework\ consistently achieves near-optimal synthesis results, particularly for bandwidth-bound large collectives, resulting in an overall average efficiency of 85.25\% relative to the ideal case.
}{1}{-2em}{-1em}

\noindent \textbf{4. \multitree.}  
We conducted a comparison between \framework\ and \multitree, a spanning tree construction-based collective synthesizer~\cite{multitree}.  
The evaluation used 2D Torus and \mesh\ topologies with $\alpha$=0.15 µs and 1/$\beta$=16 GB/s links.  
The summarized results are presented in \autoref{fig:ResultAlgorithmComparison}(a), alongside corresponding \themis\ results and the ideal upper bound.  
Initially, \framework\ demonstrated comparable performance for 1 MB \allreduce, but outperformed as the collective size increased, exhibiting an average speedup of 1.32$\times$.  
This is because, unlike \framework, which can schedule and overlap multiple chunks concurrently for maximized network resource utilization, MultiTree does not support chunk-level overlap~\cite{tto}.  
For larger collectives with multiple chunks, MultiTree cannot exploit full network bandwidth, leading to diminished results.  
Consequently, MultiTree saturated after 1 MB \allreduce, while \themis\ and \framework\ continued overlapping chunks, achieving 89.93\% and 92.15\% of theoretical efficiencies, respectively.  
Since \themis\ does not yield optimal results for asymmetric networks, for 2D Mesh, \framework\ demonstrated a notable 82.60\% efficiency.  
This underscores \framework' ability to synthesize near-optimal collectives autonomously, regardless of target topology or collective size.

\noindent \textbf{5. \ccube.}  
\ccube~\cite{ccube} manually lays out two binary trees over a DGX-1 topology and schedules two tree-based collectives concurrently.  
We modeled an equivalent DGX-1 topology with $\alpha$=0.7 µs and 1/$\beta$=25 GB/s links.  
The comparison is illustrated in \autoref{fig:ResultAlgorithmComparison}(b), where \framework\ achieved 2.86$\times$ better performance on average.  
This can be attributed to \ccube\ disabling 2 out of 6 available links per NPU to map two contention-free tree routes (as shown in Fig. 10(c) of~\cite{ccube}).  
Additionally, the remaining 4 links are not always fully utilized due to the tree-based approach for \allreduce\ (i.e., links remain idle when there are no chunks to overlap), further reducing effective network utilization.  
Compared to the ideal bound, \ccube\ achieved 32.63\% efficiency, while \framework\ reached 93.26\%, comparable to the Ring baseline at 99.61\%.

\insertFigure{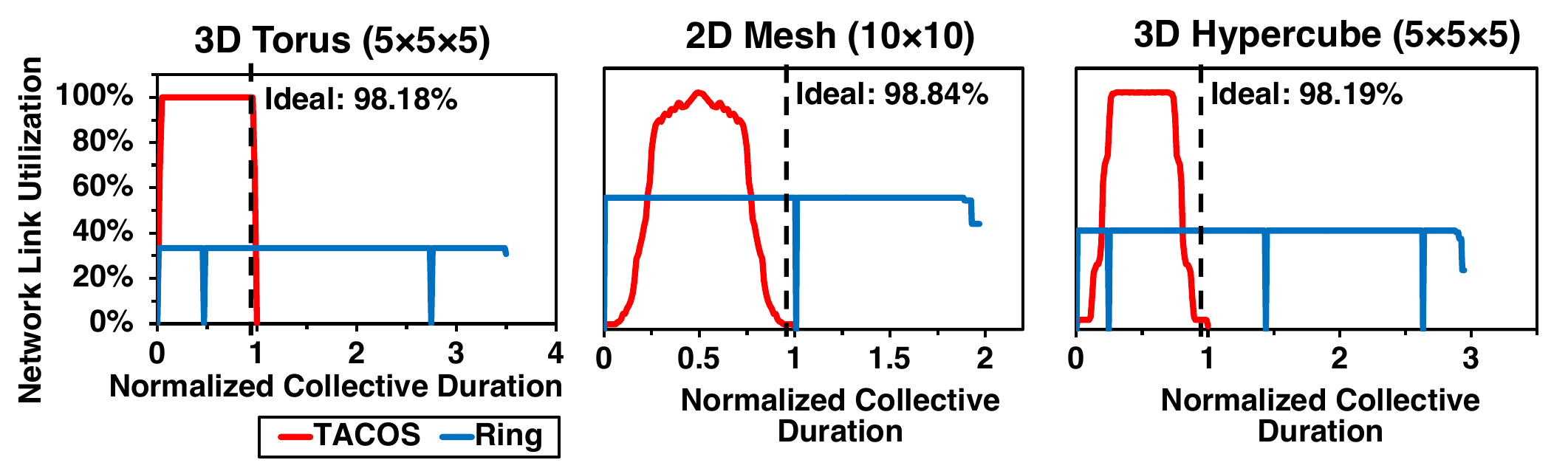}{  
Network utilization of \framework-synthesized and \ring\ algorithms during the execution of an \allreduce.  
The collective duration is normalized by \framework\ collective time.  
On average, \framework\ achieved 98.40\% efficiency compared to the theoretical ideal.  
}{1}{-2em}{-0em}

\noindent \textbf{6. Asymmetric Topologies.}  
To illustrate \framework' efficacy over asymmetric networks, as shown in \autoref{fig:ResultSynthesisUtilization}, we measured link utilization during the execution of an \allreduce\ collective over the homogeneous \torus\ (5$\times$5$\times$5), \mesh\ (10$\times$10), and \hypercube\ (5$\times$5$\times$5) topologies.  
While \mesh\ and \hypercube\ are asymmetric, \torus\ is symmetric and shown for comparison. We also included the utilization results of the \ringAlg\ algorithm.  
It is noteworthy that the symmetric \torus\ is highly suitable for collective communications, with \framework\ achieving 100\% utilization throughout the execution.  
However, the inherent asymmetry of \mesh\ and \hypercube\ topologies introduces inefficiencies, as not all chunks can depart or arrive simultaneously, causing some links to idle at the start or end of execution.  
This effect is visible at $t$=3 in \autoref{fig:MeshExampleSearchResult}.  
Nonetheless, \framework\ successfully maximizes link utilization after saturation.  
Across all scenarios, \framework\ achieved 98.40\% efficiency relative to the theoretical ideal.

\subsection{Scalability Analysis}

\insertFigure{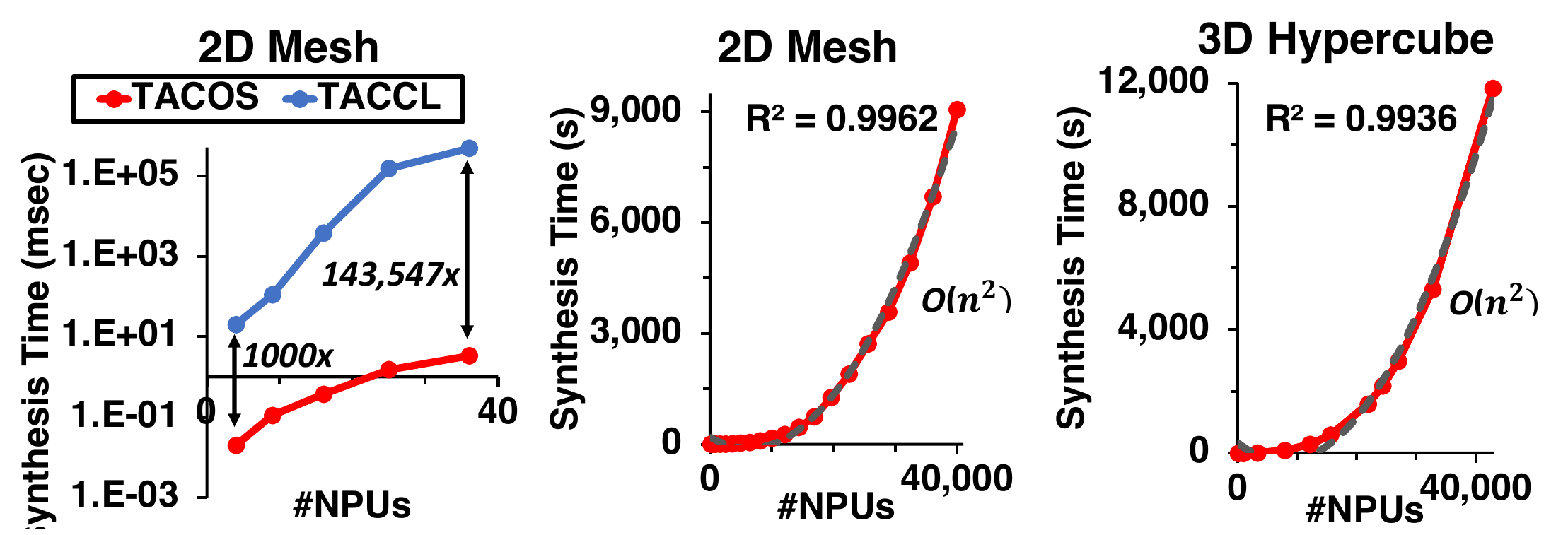}{  
Synthesis time of \framework\ and \taccl\ for various-sized homogeneous \mesh\ and \hypercube\ topologies, using \framework\ with 64 parallel threads.  
\framework\ synthesis time exhibited $O(n^2)$ complexity, where $n$ is the number of NPUs.  
}{1}{-2em}{-1em}

To demonstrate \framework' scalability, we synthesized \allreduce\ algorithms for homogeneous \mesh\ and \hypercube\ topologies with up to 43K NPUs, utilizing 64 parallel threads, and measured the synthesis time.  
The results are summarized in \autoref{fig:ResultTimeComplexity}.  
Compared to \taccl\ for up to 36 NPUs, \framework\ demonstrated significantly better scalability.  
Due to the NP-hard nature of ILP-based optimizers, the synthesis time gap between \framework\ and \taccl\ increased from $10^3$ to $10^5$ as the topology size increased by $9\times$.  

Further scalability analysis shows that \framework\ synthesized an \allreduce\ algorithm for a \mesh\ topology with 40K NPUs in 2.52 hours and for a \hypercube\ topology with 43K NPUs in 3.29 hours.  
The synthesis time scales quadratically with the number of NPUs, i.e., $O(n^2)$, where $n$ is the number of NPUs.  
This indicates that \framework' synthesis time is linear relative to the search space, which consists of $O(n)$ chunks and $\Theta(n)$ links.

\subsection{End-to-End Application}

\insertFigure{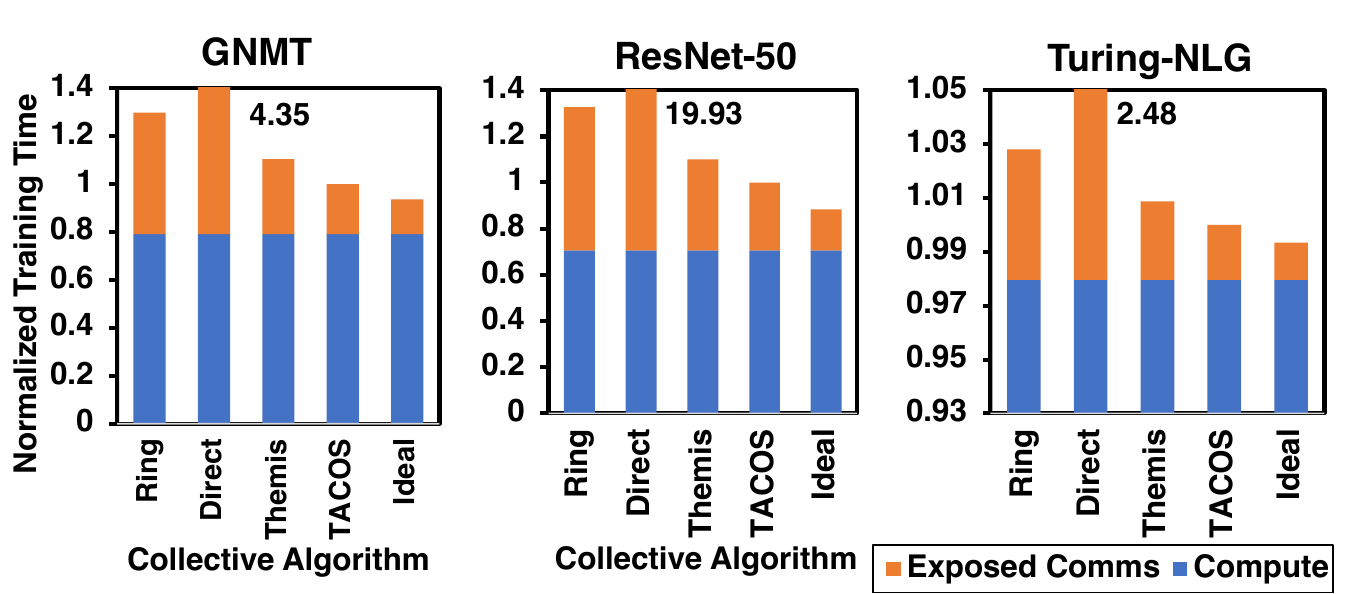}{
End-to-end training time of \gnmt, \resnet, and \tnlg.  
\gnmt\ was trained on a 64-NPU 3D-RFS system.  
\resnet\ and \tnlg\ are trained on a 3D-RFS with 32 nodes.  
All results are normalized over the corresponding \framework\ result.  
\framework\ achieved 82.38\% communication efficiency and 93.61\% end training efficiency compared to theoretical bounds.
}{1}{-2em}{-1em}

\insertFigure{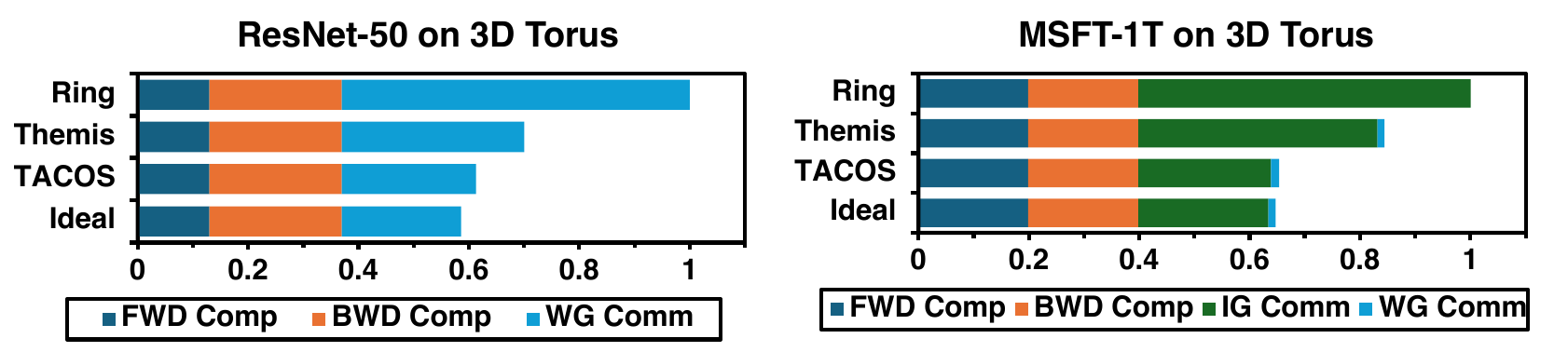}{
End-to-end training time breakdown of ResNet-50 and MSFT-1T over a 1.024-NPU, 3D Torus.  
Forward and backward pass computation, and exposed input/weight gradient communications are shown.  
All results are normalized over the Ring.  
\framework\ achieved 97.32\% efficiency over the theoretical ideal scenario.
}{1}{-2em}{-1em}

Lastly, to elucidate \framework' implications in distributed ML, we conducted an evaluation of the end-to-end training performance of \gnmt~\cite{gnmt}, \resnet~\cite{resnet}, and \tnlg~\cite{tnlg}, representative models of vision and LLM workloads.  
For models employing data parallelism, communication becomes exposed at the end of each training iteration~\cite{themis}.  
\gnmt\ was evaluated on a small, 8-node (64 NPU) \rfs\ topology, whereas \resnet\ and \tnlg\ were run over a larger 32-node (256 NPU) cluster.  
The normalized training time over \framework\ is illustrated in \autoref{fig:TrainingResult}.
Also, to model and showcase \framework' applicability for larger ML clusters, we evaluated \resnet\ and \msft~\cite{mszero} over a symmetric and homogeneous 3D Torus network with 1,024 NPUs.
The breakdown of the normalized training time is depicted in~\autoref{fig:TrainingResultNew}.

Across all workloads, leveraging \framework\ resulted in enhancements of 1.58$\times$ and 1.21$\times$ in end-to-end training performance over the baseline Ring and optimized Themis algorithms, respectively.  
Notably, compared to the theoretical upper bound, \framework\ achieved the communication efficiency of 93.17\%, thereby yielding an end-to-end efficiency of 97.32\% compared to the ideal case.

%% file: table/small_scale_result.tex
\begin{scriptsize}
\begin{table}[t]
\centering
\caption{
\allreduce\ collective time (with synthesis time in parentheses for both \framework\ and \taccl) for a multi-node 3D-RFS system with 2 to 16 nodes (16 to 128 NPUs), normalized over \framework.
On average, \framework\ achieved 75.88\% efficiency compared to the theoretical ideal.
}
\label{table:smallScaleResult}
\vspace{-0.5em}
\begin{tabular}{|c|r|r|r|r|r|r|}
\hline
\textbf{\begin{tabular}[c]{@{}c@{}}\#NPUs\\ (\#Nodes)\end{tabular}} & \multicolumn{1}{c|}{\textbf{\begin{tabular}[c]{@{}c@{}}TACOS\\ (ms)\end{tabular}}} & \multicolumn{1}{c|}{\textbf{\begin{tabular}[c]{@{}c@{}}TACCL\\ (ms)\end{tabular}}} & \multicolumn{1}{c|}{\textbf{Ring}} & \multicolumn{1}{c|}{\textbf{RHD}} & \multicolumn{1}{c|}{\textbf{Direct}} & \multicolumn{1}{c|}{\textbf{Ideal}} \\ \hline
\begin{tabular}[c]{@{}c@{}}16\\ (2)\end{tabular} & \textbf{\begin{tabular}[c]{@{}r@{}}1\\ (0.63)\end{tabular}} & \begin{tabular}[c]{@{}r@{}}2.89\\ (790)\end{tabular} & 7.14 & 5.27 & 4.04 & 1.00 \\ \hline
\begin{tabular}[c]{@{}c@{}}32\\ (4)\end{tabular} & \textbf{\begin{tabular}[c]{@{}r@{}}1\\ (8.90)\end{tabular}} & \begin{tabular}[c]{@{}r@{}}4.10\\ (2001)\end{tabular} & 5.10 & 4.42 & 7.86 & 0.72 \\ \hline
 \begin{tabular}[c]{@{}c@{}}64\\ (8)\end{tabular} & \textbf{\begin{tabular}[c]{@{}r@{}}1\\ (97.92)\end{tabular}} & \begin{tabular}[c]{@{}r@{}}4.27\\ (7016)\end{tabular} & 4.80 & 5.83 & 16.84 & 0.68 \\ \hline
\begin{tabular}[c]{@{}c@{}}128\\ (16)\end{tabular} & \textbf{\begin{tabular}[c]{@{}r@{}}1\\ (1080)\end{tabular}} & - & 4.82 & 9.85 & 36.02 & 0.68 \\ \hline
\end{tabular}
\vspace{-1em}
\end{table}
\end{scriptsize}

%% file: content/7_related_work.tex
\section{Related Work}\label{sec:related}

\subsection{Time-expanded Networks}

TEN is being leveraged in various fields that require communication or flow-based optimization.
Notable examples include vehicle traffic management~\cite{tenUse1}, logistics~\cite{tenUse2}, and traffic signal optimization~\cite{tenUse3}.
However, to the best of our knowledge, this work is the first attempt to bring TEN to distributed ML and collective communication with adequate optimization approaches.

\subsection{Collective Communications}

\noindent \textbf{Collective Algorithm Representation.}
Some works have proposed domain-specific languages for depicting collective algorithms~\cite{gc3, coconet}; however, these are oriented toward the manual design of human-crafted collective algorithms.
Other works~\cite{sccl,taccl} use custom representations.

\noindent \textbf{Co-optimization of Collective and Topology.}
Previous efforts have addressed topology optimization for specific collective algorithms~\cite{arTopologyOpt, topologyCodesign}.
Additionally, the co-optimization of collectives and network topology remains an active research domain~\cite{multitree, topoOpt}.
Given \framework's ability to perform synthesis on arbitrary topologies, it can harmonize with such endeavors.

\noindent \textbf{Collective Algorithm Manual Design.}
Designing specialized collective algorithms for specific topologies is an ongoing topic of interest. Recent examples include \tto\ for 2D Mesh~\cite{tto} and PAARD for DragonFly~\cite{dragonflyAlg}.
However, this approach requires engineering and validation efforts for each topology variant, which can be automated or augmented with synthesizers like \framework.

\subsection{Collective Algorithm Synthesizers}

\noindent \textbf{Solver-based.}
\sccl~\cite{sccl} synthesizes latency- and bandwidth-optimal collective algorithms by leveraging a satisfiability solver.
To achieve this, \sccl\ captures the design space of collective communication in linear equations.
To derive the linear relationships, \sccl\ assumes a k-synchronous collective algorithm, i.e., all NPUs executing identical transmissions clearly separated into steps and rounds.
While \sccl\ guarantees optimality, this assumption only holds for homogeneous, symmetric, single-node networks, meaning the \sccl\ approach cannot be extended beyond these scenarios.
\taccl~\cite{taccl} overcomes \sccl's limitations through an ILP approach instead of a satisfiability problem, removing the k-synchronous assumption.
However, the ILP approach requires \taccl\ to capture the entire search space in a number of linear equations.
Since network congestion effects cannot be captured in linear equations, they are completely ignored in the formulation.
Also, because the equations must be predefined, \taccl\ assumed a 2D network and constructed formulations solely based on it to model heterogeneity.
Although this assumption can be extended to multi-dimensional topologies, the number of cross-dimensional constraints would increase, further complicating the ILP search space.
Therefore, while ILP guarantees the optimal solution of the provided formulation, as the equations themselves are limited, \taccl\ does not guarantee the optimality of the synthesized collective performance.
Furthermore, \taccl\ had to assume the network is symmetrical to reduce the search space.
With all these assumptions, both \sccl\ and \taccl\ synthesis efforts still scaled to only tens of NPUs.

\noindent \textbf{Tree-based.}
\blink~\cite{blink} constructs one-to-many spanning trees from a root to execute reduction and broadcast operations, effectively performing \allreduce.
The spanning tree construction process only takes network connectivity into account, thereby yielding sub-optimal results for heterogeneous topologies.
To traverse the spanning tree upward and downward, all links are assumed to be bidirectional.
Since the spanning tree is one-to-many, the approach is naturally detrimental for many-to-many collectives such as \reducescatter or \allgather, which are required by parallelization strategies like FSDP~\cite{fsdp} or ZeRO~\cite{mszero}.
\tto~\cite{tto} optimizes Blink specifically for 2D Mesh topologies by constructing three spanning trees, all originating from edge NPUs.
Although ideal for \allreduce, \tto\ still shares the inefficiency for many-to-many collectives.
\multitree~\cite{multitree} generates height-balanced multiple spanning trees originating from all NPUs, making it suitable for many-to-many collectives.
However, \multitree\ only takes network connectivity into account, disregarding network heterogeneity.
Unlike \framework's link-chunk matching, which automatically considers multiple chunk overlaps, \multitree\ does not allow concurrent chunks to overlap.
Therefore, while both may show similar synthesis results for small, latency-critical collectives, \framework\ can achieve higher network utilization for larger collectives consisting of multiple chunks.

%% file: content/8_conclusion.tex
\section{Conclusion}

In this paper, we underscore the importance of topology-aware collective algorithms in distributed ML and emphasize the necessity of having an autonomous collective algorithm synthesizer that supports arbitrary topologies.
We introduce \framework, an automated framework for orchestrating collective algorithm synthesis.
\framework\ supports diverse networks and demonstrates near-optimal link utilization while showcasing polynomial-time scalability.

%% file: include/acknowledgment.tex
\section*{Acknowledgment}

This work was supported through awards from Intel.  
Additionally, this research is supported by the Semiconductor Research Corporation~(SRC), the SRC AIHW program, and the ACE Center for Evolvable Computing, one of the seven centers of the SRC JUMP 2.0 program.  
We also extend our sincere appreciation to Jiayi Huang and Le Qin for their assistance in helping us understand the details of MultiTree.  
We thank Ajaya Durg and Samvit Kaul for their constructive suggestions in implementing and evaluating this work.  
Finally, we greatly appreciate all the anonymous reviewers of this paper for dedicating their time and providing insightful comments to improve the quality of this work.

%% file: content/9_artifact.tex
\appendix[Artifact Appendix]

\subsection{Abstract}

We open-source and provide \framework, a topology-aware collective algorithm synthesizer, as the artifact proposed in this paper. \framework\ is a C++17-based standalone program built upon a randomized link-chunk match-making algorithm. Therefore, \framework\ supports any execution environment that can compile and execute C++17 code. \framework\ uses CMake as its build automation tool, which is its only software dependency. Since the artifact is based on a randomized algorithm, the execution results may vary, but the trend of the target metric (e.g., collective algorithm bandwidth) should remain consistent.

\subsection{Artifact Check-list (Meta-information)}

{\small
\begin{itemize}
  \item {\bf Language:} C++17
  \item {\bf Environment:} Any environment that can compile C++17
  \item {\bf Metrics:} Collective Algorithm Bandwidth (GB/s)
  \item {\bf Output:} Synthesized Collective Communication Algorithm
  \item {\bf Disk space required:} A few MBs
  \item {\bf Time to prepare workflow:} A few minutes
  \item {\bf Time to complete experiments:} Depends on the search space size (target network and collective), but only a few seconds for $O$(100s) NPUs.
  \item {\bf Publicly available?} Yes
  \item {\bf Code licenses:} MIT License
  \item {\bf Archived:} DOI:10.5281/zenodo.13325902
\end{itemize}
}

\subsection{Description}

\subsubsection{How to Access} \url{https://github.com/astra-sim/tacos}

\subsubsection{Hardware Dependencies} None.

\subsubsection{Software Dependencies} C++17 compiler and CMake.

\subsection{Installation}

The installation process is demonstrated in the \texttt{README.md} file included in the artifact (\url{https://github.com/astra-sim/tacos/blob/main/README.md}).

\noindent \textbf{1. Clone the \framework\ repository:}

\begin{itemize}[leftmargin=*]
    \item \texttt{git clone --recurse-submodules git@github.com:astra-sim/tacos.git}
    \item \texttt{cd tacos}
\end{itemize}

\noindent \textbf{2. Compile and run \framework:}

\begin{itemize}[leftmargin=*]
    \item \texttt{./tacos.sh}
\end{itemize}